\documentclass[showpacs,showkeys,nofootinbib,amsmath,amssymb]{revtex4}

\usepackage{mathrsfs}
\usepackage{graphicx,epsfig}
\usepackage{psfrag}
\usepackage{enumerate}
\usepackage{citesort}
\usepackage{hyperref}

\newcommand{\ttau}{\tilde{\tau}}
\newcommand{\bea}{\begin{eqnarray}}
\newcommand{\beal}[1]{\begin{eqnarray}\label{#1}}
\newcommand{\eea}{\end{eqnarray}} 
\newcommand{\be}{\begin{equation}} 
\newcommand{\bel}[1]{\begin{equation}\label{#1}}
\newcommand{\ee}{\end{equation}} 
\newcommand{\rf}[1]{(\ref{#1})}
\newcommand{\nn}{\nonumber}
\newcommand{\bit}{\begin{itemize}}
\newcommand{\eit}{\end{itemize}}
\newcommand{\ben}{\begin{enumerate}}
\newcommand{\een}{\end{enumerate}}

\newcommand{\npd}{\nabla_\perp}
\newcommand{\npu}{\nabla^\perp}
\newcommand{\sn}{s}
\newcommand{\mt}{\left(\nabla\cdot u\right)}

\newcommand{\tq}{\tilde{q}}

\newcommand{\tom}{\tilde{\omega}}

\newcommand{\Lie}{\mathcal{L}}

\newcommand{\tl}{\theta_{(\ell)}}
\newcommand{\tn}{\theta_{(n)}}
\newcommand{\cV}{\mathcal{V}}
\newcommand{\vS}{\sqrt{\tilde{q}}}
\newcommand{\norm}{| \mspace{-2mu} |}

\def\half{\frac{1}{2}}
\def\third{\frac{1}{3}}

\def\thetan{\theta_{(n)}}
\def\thetal{\theta_{(\ell)}}

\def\alp{\leavevmode\ifmmode {\alpha^\prime} \else ${\alpha^\prime}$ \fi}

\begin{document}

\title{Black brane entropy and hydrodynamics: \\
the boost-invariant case}

\author{Ivan Booth}
\email[]{ibooth@mun.ca}
\affiliation{Department of Mathematics and Statistics\\
Memorial University of Newfoundland\\
St. John's, Newfoundland and Labrador, A1C 5S7, Canada}
\author{Michal P. Heller}
\email[]{michal.heller@uj.edu.pl}
\affiliation{Institute of Physics, Jagiellonian University\\
Reymonta 4, 30-059 Cracow, Poland}
\author{Michal Spalinski}
\email[]{mspal@fuw.edu.pl}
\affiliation{So\l tan Institute for Nuclear Studies, ul. Ho\.za 69, 
00-681 Warsaw, Poland \\
and Physics Department, University of Bialystok, 15-424 Bialystok, Poland. }

\begin{abstract}
The framework of slowly evolving horizons is generalized to the case of black
branes in asymptotically anti-de Sitter spaces in arbitrary dimensions. The
results are used to analyze the behavior of both event and apparent horizons in
the gravity dual to boost-invariant flow.  These considerations are motivated by
the fact that at second order in the gradient expansion the hydrodynamic entropy
current in the dual Yang-Mills theory appears to contain an ambiguity.  This
ambiguity, in the case of boost-invariant flow, is linked with a similar freedom
on the gravity side. This leads to a phenomenological definition of the entropy
of black branes. Some insights on fluid/gravity duality and the definition of
entropy in a time-dependent setting are elucidated.
\end{abstract}

\pacs{11.25.Tq,
%Gauge/gravity duality
04.50.Gh,
%Higher-dimensional black holes
04.20.Gz.
%Causal structure
}

\keywords{Gauge/gravity duality, Black Holes, Quasilocal horizons.}

\maketitle

\section{Introduction}

One of the most important and intriguing recent developments in
theoretical physics is the AdS/CFT correspondence.  An example of holographic
gauge/string theory duality \cite{adscft} asserts the complete physical
equivalence between particular string vacua and ordinary quantum field
theories. In certain regimes it relates strongly coupled field theories to
weakly curved string theory, which in the leading order reduces to
supergravity. A very interesting class of applications involves systems
at finite temperature. The holographic duality relates thermodynamic notions in
quantum field theory to black hole mechanics in the bulk description. 
Even more fascinating are non-equilibrium phenomena which cannot be described by
thermodynamics or static black hole solutions. There is very strong motivation
for such studies on both sides of AdS/CFT duality.

While there are several frameworks that provide fundamental explanations for the entropy of a 
stationary black hole, the entropy of dynamical black holes is an even more difficult problem. 
This issue, in view of AdS/CFT
duality, is directly connected with the physics of non-equilibrium processes in
strongly coupled quantum field theory. Progress in this area will have
important bearing on problems in black hole as well as particle
physics. 

The past few years have seen tremendous activity applying gauge/gravity duality
to study gauge theories beyond the perturbation series. One obvious application
area where such a tool is sorely needed is the investigation of non-perturbative
dynamical QCD plasma. The case of $\mathcal{N}=4$ super Yang-Mills (SYM) theory has
been the focus of attention in this context, since its holographic
representation in terms of string theory on anti-de Sitter spacetimes is the
best understood example of gauge/gravity duality\footnote{Many of the results
  hold 
  also for sectors of other large N$_{c}$ strongly coupled gauge theories which
  have gravity dual described by the Einstein-Hilbert action with negative
  cosmological constant. This is another hint pointing towards possible
  (limited) applicability of AdS/CFT results to the real world.}.  Furthermore,
it appears that the significant differences between this theory and real-world QCD do not play a
major role in a particular  range of temperatures relevant to the heavy ion
experiments currently in progress at RHIC and soon to start at the LHC.  
Thus the studies of Yang-Mills plasma
at finite temperature using tools provided by string theory (such as
\cite{Janik:2005zt,Nakamura:2006ih,Janik:2006ft,Heller:2007qt,Benincasa:2007tp,Bhattacharyya:2008jc,Bhattacharyya:2008xc,VanRaamsdonk:2008fp}) are currently of great practical importance and attract a lot of attention from
the heavy ion community.

It is now fairly well established that there is a regime where the dynamics of
quark-gluon plasma is well described by relativistic hydrodynamics. The
hydrodynamic description itself is well understood despite the fact that the
underlying quantum field theory is strongly coupled. It is based on the idea of
an expansion in the number of gradients, much in the spirit of effective field
theory. One spectacular application of AdS/CFT in recent years was to show that
in the case of $\mathcal{N}=4$ SYM a hydrodynamic description can be derived by
considering time-dependent black-brane solutions in AdS spacetime (particularly useful review is \cite{Rangamani:2009xk}). Further
applications to quark-gluon plasma require a better understanding of the duality
at finite temperature and in a dynamical setting.

A byproduct of the hydrodynamic construction is the notion of entropy
current. Recent advances in relativistic conformal hydrodynamics show that a
phenomenologically introduced hydrodynamic entropy current consturucted order by
order in the gradient expansion involves an ambiguity
\cite{Bhattacharyya:2008xc,Romatschke:2009kr}. At second order a 4-parameter
family of currents\footnote{In general there might be a 5-parameter ambiguity, but one
  of these parameters is connected with parity-odd effects. This article
  considers only the parity-even case.}  was identified in
\cite{Bhattacharyya:2008xc}. Subsequently some important considerations on this
topic have appeared in the literature \cite{Romatschke:2009kr}, which suggest
that it may be possible to reduce this freedom to just one parameter. It seems
however 
that there is a real ambiguity in the hydrodynamic entropy current and in the
light of AdS/CFT duality it is natural to ask if one can understand its origins
on the gravity side. One of the goals of the present study is to address this
issue.

At the root of this question lies the identification of the area increase
theorems in general relativity with the second law of thermodynamics. 
Thus if one hopes to understand this ambiguity on the gravity side, the first step is 
to carefully examine and understand how horizon areas increase. In doing
this one may build on the experience gained in the study of standard black
holes in asymptotically flat four-dimensional spacetime. Generalizing
these results to five-dimensional anti-de Sitter spacetime is not difficult, 
but the deep questions encountered earlier remain. 

The most ``conservative'' definition of entropy identifies it with the
area of a spatial section of the event horizon. This notion has the drawback
of being a teleological (and thus global) concept. In the context of AdS/CFT
duality this leads to acausality in the field theory
\cite{Chesler:2008hg,Figueras:2009iu}.  In classical general relativity, the 
problems raised by the global character of the event horizon have lead to alternative, 
quasilocal notions of black holes. These include trapping \cite{hayward}, isolated \cite{IsoRef}, 
and dynamical \cite{DynRef} horizons. These programmes are closely related to each 
other and motivated by historical ideas about trapped surfaces \cite{penrose} and 
apparent horizons \cite{Hawking:1973uf}. 

These quasilocal horizons have a problem of their own, namely non-uniqueness:
unlike event horizons, in dynamical black-hole spacetimes, there are many
possible time-evolved apparent (or trapping or dynamical) horizons. In the case
of apparent horizons, each foliation of the spacetime will give rise to a
different time-evolved horizon \cite{Hawking:1973uf} while trapping/dynamical
horizons are also subject to deformations (see, for example, \cite{GregAbhay}
\cite{Booth:2006bn}). Though this is generally thought to be a bad thing, in the
context of AdS/CFT it is natural to suspect that the ambiguity of the
hydrodynamic entropy current may be related to ambiguities of this type.

In this paper it is shown that the non-uniqueness of apparent horizons is not
the source of the ambiguity in the entropy current. Instead, the already
existing uncertainty as to whether the event or apparent horizon determines the
entropy is embraced and a more general approach is advocated which mimics the
phenomenological construction of the boundary entropy current. In the spirit of
the membrane paradigm \cite{membrane} we consider ``horizons'' that are made up from families of
(not necessarily trapped) codimension-two surfaces that satisfy properties such
as area increase, asymptoting to the correct equilibrium limits, and being
``almost'' apparent horizons.  In general such families are hard to deal with,
but the situation becomes much more manageable close to equilibrium. We consider
spacetimes in which a small parameter can be used to formulate a perturbative
expansion around equilibrium (this does not necessarily imply a hydrodynamic
description \cite{Chesler:2008hg}). Such a formalism -- slowly evolving geometry
-- has been developed in the context of regular four-dimensional black holes
\cite{Booth:2003ji,Booth:2006bn,Kavanagh:2006qe} and it is adapted to the case
of black brane mechanics as well as generalized to the case of
not-quite-apparent horizons. This formalism also makes it possible to see the
conditions under which the first law of thermodynamics can be formulated in a
dynamical setting.

It is conjectured here that the ambiguity in the above definition of black brane
entropy corresponds to the known ambiguity of the hydrodynamic entropy current
in the appropriate regime on the gravity side. Verifying this claim in the
general case is beyond the scope of this paper, which explores this question in
a special, highly symmetric case -- the gravity dual to Bjorken
(boost-invariant) flow.  Besides being tractable, this geometry has the
important feature that there is a unique apparent horizon consistent with
symmetries of boundary flow. This means that the ambiguity in the hydrodynamic
entropy current in case of Bjorken flow cannot be interpreted as a consequence
of slicing dependence.

Boost-invariant flow is one of the simplest, yet phenomenologically interesting
examples of boundary dynamics.  It is a one-dimensional expansion mimicking the
dynamics of plasma created in the heavy ion collision with an additional
assumption of boost-invariance along the collision axis. This is based on
Bjorken's observation \cite{Bjorken} that multiplicity spectra when expressed in
proper time and rapidity variables are approximately independent of rapidity in
the mid-rapidity region. While this approximation may be somewhat rough, it
provides a dramatic simplification. After taking into account all the symmetries
in this problem, it turns out that all the physical quantities depend on single
variable - the proper time. In the pioneering work \cite{Janik:2005zt} Janik and
Peschanski considered the gravity dual to boost-invariant flow in the regime of
large proper time and showed that regularity of the bulk geometry forces perfect
fluid hydrodynamics on the boundary. Subsequent developments included
calculating subleading corrections to the large proper time geometry, which
correspond to dissipative terms in the boundary energy-momentum tensor
\cite{Nakamura:2006ih,Janik:2006ft} (see \cite{Heller:2008fg} for a useful review). Understanding the gravity dual to Bjorken flow within the framework of fluid-gravity duality
\cite{Heller:2009zz,Kinoshita:2008dq} makes it possible to address issues of
black brane mechanics in this case\footnote{It is critical to have a smooth
  description of the horizon region.}.

One can also look at this paper from another perspective. In understanding any
mathematical formalism, it is very useful to have concrete examples with which
to work. There are very few exact (or even perturbative) dynamical black holes
solutions on which to test ideas such as those about slowly evolving
horizons. Most of the already known examples were examined in
\cite{Kavanagh:2006qe}. These included spherically symmetric spacetimes where
the expansion was driven by matter flows and perturbations of Schwarzschild
where the expansion was driven by shears (gravitational waves) -- though
unfortunately in this case the order of the perturbation  was such that one could
see the shears driving the expansion but not the expansion itself. Then, from this point of view, the current
work presents a new non-numerical example of a near-equilibrium horizon and it
is the first one for which a shear driven expansion can actually be directly
observed.  This allows equations to be checked in previously inaccessible
regimes and also helps to build a better understanding of the formalism.

The structure of the paper is the following. Section two reviews the various
definitions of black hole and then develops black hole mechanics in AdS$_{n+1}$
spacetime, generalizing \cite{Booth:2003ji,Booth:2006bn} and keeping things as
general as possible with a view to further applications of these results.  The
third section gives a brief introduction to both field theory and gravity
aspects of the boost-invariant flow and the fourth section applies methods
developed in section two to the boost-invariant situation. The fifth section
contains the analysis of the freedom in the definition of entropy on both sides
of the AdS/CFT correspondence in the case of Bjorken flow and shows that there
is precise match. 
The final section
offers conclusions and signals some directions for further research.

\section{Black brane mechanics\label{BBmech}}
\label{blabrame}

\subsection[A brief review of black holes and horizons]{A brief review of black holes and horizons}
\label{HorizonIntro}

We begin with a review of three ways of thinking about black holes. Since our
main interest in this paper are the dynamically evolving black branes dual to
Bjorken flows, we will focus in particular on dynamical black holes. While the
various definitions of black holes generally agree for stationary black holes, they
diverge away from equilibrium. For example, they identify different surfaces in
spacetime as the boundary of the black hole region and these surfaces have
different surface areas. This will be significant in later sections when we
compare the thermodynamics on the CFT side with the corresponding black hole
mechanics on the gravity side.  Then, as noted in the introduction, different
surface areas would suggest different values of entropy.

\subsubsection*{Event horizons: causally defined black holes}

It is well known that causally defined black holes and event horizons are not
locally defined. Paraphrasing, a classically-defined black hole
\cite{Hawking:1973uf} is a region of spacetime from which nothing can ever
escape. At first glance this seems like a perfectly reasonable definition,
however a little consideration quickly turns up problems and these arise from
the concepts of ``ever'' and ``escape''. To identify a black hole region one
must essentially sit at infinity and wait forever to make sure that all escaping
signals are identified and further that those that initially look like they
might escape really do make it to infinity. Equivalently (but more rigorously)
the black hole region is the complement of the causal past of $\mathscr{I}^+$
(future null infinity).  The boundary of the black hole region is the congruence
of null geodesics known as the \emph{event horizon}.

%\EPSFIGURE{EHFig.eps}{A schematic demonstrating the non-local nature of
%  event horizon evolution for a spherically symmetric spacetime with the
%  angular dimensions suppressed. Horizontal location measures the radius of
%  the associated spherical shell while time is (roughly) vertical. The
%  shaded gray region represents infalling null dust. \label{EHfig}} 

\begin{figure}
\caption{\label{EHfig} A schematic demonstrating the non-local nature of
event horizon evolution for a spherically symmetric spacetime with the
angular dimensions suppressed. Horizontal location measures the radius of
the associated spherical shell while time is (roughly) vertical. The
shaded gray region represents infalling null dust.}
\includegraphics{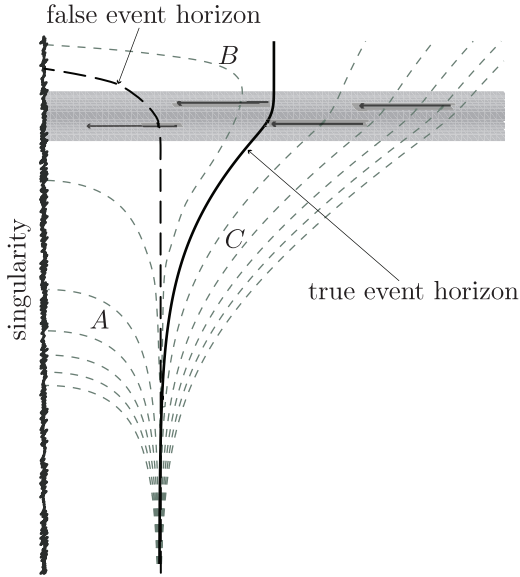}
\end{figure}

To better understand this definition and its associated peculiarities consider
Figure \ref{EHfig} which represents the evolution of an initially isolated
Schwarzschild black hole which is later irradiated by an infalling shell of
null dust (see \cite{BoothMartin} for more discussion of the Vaidya
spacetimes used to generate this example). In the diagram the inward null
direction is horizontal while the outward null directions are tangent to
the light gray dashed lines. Thus in general the future of any point is
``up-and-to-the-left'' with the light cones pointing more and more towards
the singularity as one approaches it.

In this case one simply tracks the evolution of radial null geodesics to
find the event horizon -- if they fall into the singularity they are inside
the black hole but if they are still heading outwards after the shell of
matter passes then they are deemed to escape to infinity.  Intially,
without any knowledge of the future arrival of the shell of matter, one
would guess that the heavily dashed line at $r=2M_o$ ($M_o$ the mass
parameter in the initial Schwarzschild spacetime) would be the event
horizon: null geodesics inside it (labelled A in the diagram) clearly
head inwards towards the singularity while $r=2M_o$ itself keeps a constant area. 
However, with the arrival of the
matter shell one finds that this false event horizon also terminates at
$r=0$ as do apparently escaping geodesics like B. It then turns out that
the true event horizon was a congruence of geodesics that initially
appeared to be escaping but was later ``almost caught'' by the increased
gravitational field arriving with the shell. After the passage of the shell
these become stationary $r=2(M_o + \Delta M)$, where $M_o + \Delta M$ is
the new Schwarzschild mass.

This example nicely demonstrates the consequences of a causal structure
definition of black holes.  While the evolution of any congruence of null
geodesics is certainly causal, the identification of the set corresponding to
the event horizon depends on events in the far future.  As a result
non-omniscient observers cannot precisely locate them. Further even if
identified they will evolve in non-intuitive ways: in the example, the arrival
of matter didn't cause the event horizon to expand but instead curtailed an
expansion that began earlier (in a sense in anticipation of the mass increase).

\subsubsection*{Apparent horizons: geometrically defined black holes}

Given these peculiarities, it is often more useful to turn to alternative
definitions of black holes and their boundaries which leave aside the causal
structure of spacetime and instead focus on the strong gravitational fields
characterized by the existence of \emph{trapped surfaces}. For regular
four-dimensional astrophysical black holes, trapped surfaces are closed and
spacelike two-surfaces which have the property that all families of null
geodesics that intersect them orthogonally must converge into the future. To
understand this intuitively, consider a transparent spherical shell that is
covered with light bulbs and sitting in empty space.  Then if the bulbs are
quickly turned on and then off again, two spherical light fronts will be
generated -- an outwards moving one that expands in area and an inwards moving
one that contracts. By contrast, if the shell is transported so that it lies
inside a Schwarzschild black hole, concentric with the horizon and enclosing the
singularity, then both light fronts will fall towards the centre of the black
hole and contract in area; again consider Figure \ref{EHfig} where both outward
and inward falling congruences contract in area. This is the canonical example
of a trapped surface.

More mathematically, if $\ell^a$ and $n^a$ are respectively the outward and
inward pointing null normals to a two-surface $S$ then one can write
\bea
\theta_{(\ell)}  < 0  \; \text{and} \; \theta_{(n)} < 0 \, , 
\eea
where $\theta_{(\ell)}$ and $\theta_{(n)}$ are the \emph{expansions} of the
null normals and are geometrically analogous to  
(traces of) extrinsic curvatures 
\bea
\theta_{(\ell)}  = \tq^{ab} \nabla_a \ell_b   \; \text{and} \; \theta_{(n)}
= \tq^{ab} \nabla_a n_b \,  \label{tLtN}
\eea
with 
\bea
\tq_{ab} = g_{ab} + \ell_a n_b + n_a \ell_b  \label{SurfMet}
\eea
being the induced (spacelike) metric on the two-dimensional surface. Alternatively given 
outward and inward congruences of null geodesics which have tangents
$\ell^a$ and $n^a$ on $S$ one can show that 
\bea
\sqrt{\tq} \tl = \frac{1}{2} \Lie_\ell \sqrt{\tq} \; \text{and} \;
\sqrt{\tq} \tn = \frac{1}{2} \Lie_n \sqrt{\tq} \, ,  
\eea
where $\sqrt{\tq}$ is the area element on $S$ and $\Lie$ indicates is the
Lie derivative operator. Then it is clear that the sign 
of the expansion determines whether the congruence is expanding or
contracting.

More generally, given the energy conditions the mere existence of a trapped surface in an 
asymptotically flat spacetime implies both: 
1) a singularity somewhere in its centre (future) \cite{penrose} and 2)
that it is necessarily contained in a causal black hole and  
so event horizon \cite{Hawking:1973uf}. Indeed, for the standard Kerr family of
(stationary) black hole solutions, the set of all points contained 
on some trapped surface coincides exactly with the black hole region. 
Thus it is not unreasonable to consider the existence of trapped surfaces as being the key 
characterizing feature of a black hole region and this is the basis of the
alternative definitions of black holes.  

%\EPSFIGURE{ApparentHorizon.eps}{An ``instant'' $\Sigma_t$ along with
%  some of its trapped surfaces (small black circles), the associated
%  trapped region (dark gray) and the apparent horizon (thick dashed
%  line). } 

\begin{figure}
\caption{An ``instant'' $\Sigma_t$ along with some of its trapped surfaces (small black circles), the associated
trapped region (dark gray) and the apparent horizon (thick dashed line).}
\includegraphics{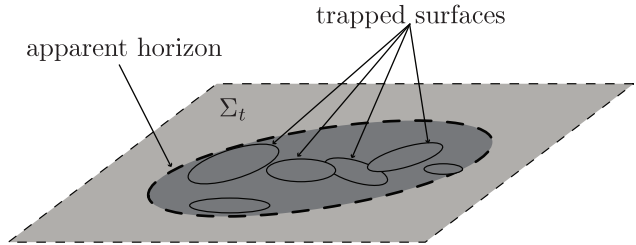}
\end{figure}

The original such definition was the \emph{apparent
  horizon}. This begins with the foliation of a spacetime into spacelike
hypersurfaces -- essentially instants in time. Then at a given instant, the
\emph{trapped region} is the union of all the trapped surfaces contained in the hypersurface
and the boundary of that region is the \emph{apparent horizon}. It can
be shown \cite{Hawking:1973uf}, that on the apparent horizon $\tl = 0$ and $\tn
< 0$. More generally, any such surface satisfying these conditions is
referred to as \emph{marginally trapped}\footnote{Unfortunately there is
  some inconsistency in the literature about the usage of this phrase. That said, our usage here is probably the most common one -- for more
details on usage see \cite{Booth:2005qc}. }. 

This observation that apparent horizons are marginally trapped then
motivates the various modern notions of quasilocal horizons such as
\emph{trapping} \cite{hayward}, \emph{isolated} \cite{IsoRef}, and
\emph{dynamical} horizons \cite{DynRef} (or see review articles such as
 \cite{Booth:2005qc} and \cite{AbhayBadri}). Though there are significant
technicalities, the idea is that the identification of marginally trapped
surfaces, which under arbitrarily small deformations become fully trapped,
is sufficient to signal the presence of a black hole boundary -- even
without going through the process of foliating the spacetime and finding
apparent horizons on each slice. In fact, this idea is so pervasive that in
numerical relativity \cite{thomas}, the term ``apparent
horizon'' has been co-opted to refer to the outermost surface $\tl
=0$ on a give slice of spacetime. The study of these ideas is an active and developing area of research
with a fairly complicated system of nomenclature but for the purposes of
this paper, quasilocal black hole (or brane) horizons in a
$(n+1)$-spacetime dimensions will be understood as $n$-dimensional hypersurfaces that are
foliated by $(n-1)$-dimensional marginally trapped spacelike surfaces. The
requirement that there be fully-trapped surfaces ``just inside'' the
horizon can be mathematically written as: 
\bea 
\Lie_n \tl < 0 \, , 
\eea
where in this case $n^a$ is any extension of the null normal $n^a$ into a
neighbourhood of the putative horizon \cite{Booth:2006bn}. In what follows we will usually 
adopt Hayward's nomenclature and refer to such structures as future outer trapping horizons (FOTH) \cite{hayward}. 
Occasionally however we will refer to time-evolved apparent horizons (which are examples of FOTHs) or dynamical horizons \cite{DynRef}
which are almost equivalent. 

%\FIGURE{
%\epsfig{file=HorizonsDouble.eps} 
%\caption{A simulation similar to that of Figure
%  \ref{EHfig} though this time two distinct shells fall into the black
%  hole.  Both the apparent and event horizons are plotted.} \label{doublehor}}

\begin{figure}
\caption{\label{doublehor} A simulation similar to that of Figure \ref{EHfig} though this time two distinct shells fall into the black hole.  Both the apparent and event horizons are plotted.}
\includegraphics{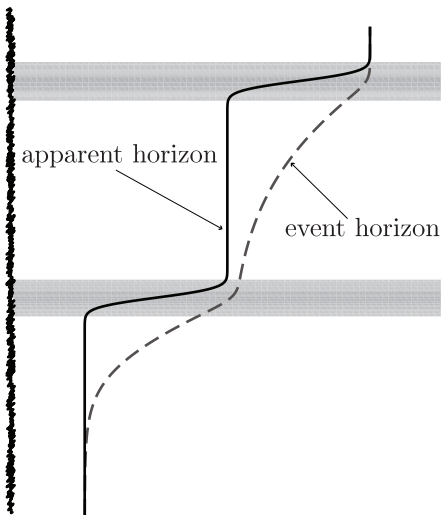}
\end{figure}

Mathematically the properties of these
surfaces (such as existence, uniqueness, and evolution) may be studied
using standard techniques from differential geometry. From a physical
perspective however, the key advantages of these generalized apparent
horizons include: 1) they are defined by the existence of strong
gravitational fields, 2) as for fully trapped surfaces their existence is
sufficient to imply the existence of singularities and event horizons, 3)
they may be identified without reference to the far future (if one thinks
of them as time-evolved apparent horizons), and 4) their evolution is
similarly local. The local nature of this evolution is demonstrated in
Figure \ref{doublehor}. In that case, the expansion of the event horizon
continues to occur in anticipation of the arrival of infalling matter, with the actual 
arrival of the matter slowing or ending that expansion. By contrast the
apparent horizon evolves in the expected way in response to the infalling
matter -- it expands in and only in the presence of actual matter crossing
the horizon.

On the downside, it is well-known that quasilocal horizons are not uniquely
defined. For classically defined apparent horizons this is easily be seen. Given
a foliation of spacetime we can define a time-evolved apparent horizon
$\triangle$ as the union of the apparent horizons on each surface. Then, it is
clear that different foliations will sample different subsets of all the
possible trapped surfaces. Thus, different foliations will define different
$\triangle$. In the extreme case it is known that certain slicings of
Schwarzschild spacetime contain no trapped surfaces at all and so no apparent
horizon \cite{waldiyer}. We will return to this lack of uniqueness in later
sections.

\subsubsection*{The membrane paradigm: a physical approach to black holes}

A third way of looking at black holes focuses not on causality or geometry but
rather on how black holes interact with their environment. By definition event
horizons cannot directly affect their surroundings (they are not in causal
contact with any point outside themselves) and neither can apparent horizons
(they are contained within event horizons).  All that either can do is impose
restrictions on the behaviour of surfaces ``near'' the horizons that are in
causal contact with the outside. The \emph{membrane paradigm} \cite{membrane}
studies the physics of $n$-dimensional timelike hypersurfaces ``just outside''
the event horizon.  For astrophysical purposes, it turns out that one can view
these as the time evolutions of $(n-1)$-dimensional spacelike viscous fluid
surfaces which carry energy, angular momentum and entropy and can exchange these
with their surroundings. The details of this formalism are not important here
but it is important to understand that for some purposes surfaces besides the
usual horizons can usefully be thought of as physically characterizing black
holes.

\subsection{The geometry of $n$-tubes}
\label{geometry}

%\EPSFIGURE{Cdef.eps}{A schematic of an $n$-tube $\triangle$
%  with compact foliation surfaces $S_\lambda$ along with the outward and inward
%  pointing null normals to those surfaces. $\mathcal{V}^a$ is the
%  future-pointing tangent to $\triangle$ that is simultaneously normal to the
%  $S_\lambda$. 
%  \label{Cdef}} 

\begin{figure}
\caption{\label{Cdef} A schematic of an $n$-tube $\triangle$
with compact foliation surfaces $S_\lambda$ along with the outward and inward
pointing null normals to those surfaces. $\mathcal{V}^a$ is the
future-pointing tangent to $\triangle$ that is simultaneously normal to the $S_\lambda$. }
\includegraphics{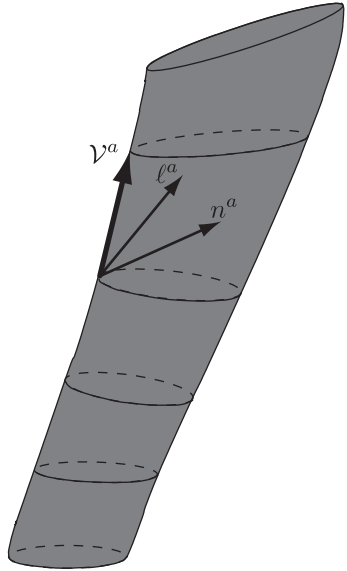}
\end{figure}

With the motivation of the last section in mind,
the mathematics of \emph{n-tubes}, a class of geometric structures that includes
both FOTHs and event horizons as well as the time-like hypersurfaces of the membrane 
paradigm, will be examined here. In an
$(n+1)$-dimensional spacetime $n$-tubes are $n$-dimensional surfaces which can
be foliated by $(n-1)$-dimensional spacelike surfaces $S_\lambda$.  The term
``tube" comes from $(3+1)$-dimensions where the $S_\lambda$ for horizons are
compact and generally diffeomorphic to $S^2$ (Figure \ref{Cdef}), however this
nomenclature will be kept even for black branes
where the $S_\lambda$ are diffeomorphic to $\mathbb{R}^3$ (or in the case of
$AdS_{n+1}$ with $\mathbb{R}^{n-1}$) and so certainly not compact.

Foliated event horizons, time-evolved apparent horizons and membranes are clearly
examples of such structures. Event horizons are $n$-tubes of null signature
which have the correct causal properties as discussed in the previous section. 
Time-evolved
apparent horizons are FOTHs whose $S_\lambda$ are
the apparent horizons found in individual space-time slices. As will be clear in
moment, these are either null (if isolated and in equilibrium) or spacelike (if
dynamical and expanding). Finally in the membrane paradigm the $\triangle$ is a timelike surface
and the $S_\lambda$ can be thought of as the evolving configurations of a fluid. 

To begin, consider the basic geometry of $n$-tubes and in particular focus on
the spacelike $S_\lambda$. First, the co-dimension of the  
$S_\lambda$ is two and the normal space has Minkowski signature so, as in the brief review, 
one can always find null normals $\ell_a$ and $n_a$ which span that normal space. $\ell^a$ is taken to be 
outward-pointing (and so is tangent to $\triangle$ if it is null) while $ n^a$
points inwards towards the singularity; again see Figure \ref{Cdef}.  
For convenience the normals are usually cross-normalized so that $\ell \cdot n =
-1$ which leaves a single scaling degree of 
freedom in their definition
\bea
\ell^a \rightarrow f \ell^a  \; \mbox{and} \; n^a \rightarrow \frac{1}{f} n^a
\eea
for any positive function $f$. 

The intrinsic geometry of the $S_\lambda$ is defined by the induced metric $\tilde{q}_{ab}$ (\ref{SurfMet})
while the extrinsic geometry is characterized by the derivatives of the normal vectors in the directions tangent to $S_\lambda$. 
These include the \emph{extrinsic curvatures}:
\bea
k^{(\ell)}_{ab} = \tq_a^c \tq_b^d \nabla_c \ell_d \; \mbox{and} \; k^{(n)}_{ab} = \tq_a^c \tq_b^d \nabla_c n_d \, ,
\eea
which decompose into traces $\tl$ and $\tn$ as well as trace-free parts $\sigma^{(\ell)}_{ab}$ and $\sigma^{(n)}_{ab}$:
\bea
k^{(\ell)}_{ab} = \frac{1}{(n-1)} \tl \tq_{ab} + \sigma^{(\ell)}_{ab}  \; \mbox{and} \; 
k^{(n)}_{ab} = \frac{1}{(n-1)} \tn \tq_{ab} + \sigma^{(n)}_{ab} \, . \label{kDecomp}
\eea
Physically these are respectively the expansions and shears of congruences of null curves which have tangents 
$\ell^a$ and $n^a$ as they intersect $S_\lambda$. 

The rest of the extrinsic geometry of the $S_\lambda$ is described by the connection on the normal bundle:
\bea
\tom_a = - \tq_a^b n_c \nabla_b \ell^c \, .  
\eea
The geometric information contained in $\tom_a$ is somewhat obscured by its gauge-dependence on the scaling $f$:
\bea
\ell \rightarrow f \ell \; \mbox{and} \; n \rightarrow \frac{1}{f} n \;
\Longrightarrow \; \tom_a \rightarrow \tom_a + d_a \ln f \, , \label{gaugetom} 
\eea
where $d_a$ is the $(n-1)$-dimensional gradient operator on $S_\lambda$. 
In the usual way for gauge potentials, the invariant information is contained in its curvature: in this case
this is the curvature of the normal bundle
\bea
\Omega_{ab} = d_a \tilde{\omega}_b - d_b \tilde{\omega}_a \, .  
\eea

These normal-bundle quantities will arise in the following discussions
and physically are closely related to angular momentum  
(see for example the extensive discussion in \cite{Booth:2006bn}). That said, in
the actual application of this work to our boost-invariant 
spacetimes, $\tom_a$ will vanish thanks to the symmetry of the black-branes. 

The next natural step is to consider how the $S_\lambda$ fit together to form
$\triangle$. Let $\mathcal{V}^a$ denote the evolution vector field tangent to  
$\triangle$ that maps leaves of the foliation into each other ($\Lie_{\cV}
\lambda = 1$) and is normal to each of the $S_\lambda$.  
Then $f$ can always be chosen so that 
\bea
\mathcal{V}^a = \ell^a - C n^a \, \label{cV}
\eea
for some function $C$. This ties the scaling of the null vectors to the
foliation and the freedom is reduced to that in the  
foliation labelling: that is $f = f(\lambda)$. In particular, from
(\ref{gaugetom}) it is straightforward to see that this fixes $\tom_a$,  
removing its gauge dependence.  

Clearly thanks to the normalization $\ell \cdot n = -1$, $C > 0 \Leftrightarrow
\triangle$ is spacelike, while $C = 0 \Leftrightarrow \triangle$ is null, and $C
< 0 \Leftrightarrow \triangle$ is timelike. Further the ``time"-rate of change
of the area element\footnote{Quotation marks 
are used around the word time since if $\mathcal{V}^a$ is spacelike then this is
a coordinate rather than physical notion of time. } is 
\bea
\Lie_{\cV} \sqrt{\tq} = \sqrt{\tq} (\tl - C \tn) \, . \label{AreaExp}
\eea
Thus, for apparent horizons with $\tl = 0$ and $\tn < 0$, $C$ also characterizes
the evolution of the horizons and is often  
referred to as the \emph{expansion parameter}. Specifically in such cases
\begin{equation}
\begin{array}{cclcl}
C < 0  &   \Leftrightarrow & \sqrt{\tq} \text{ is decreasing} & \Leftrightarrow & \cV^a \text{ is timelike,}  \\
C = 0  &   \Leftrightarrow &  \sqrt{\tq} \text{ is unchanging} & \Leftrightarrow & \cV^a \text{ is null,} \\
C > 0  &   \Leftrightarrow & \sqrt{\tq} \text{ is increasing} & \Leftrightarrow & \cV^a \text{ is spacelike.}
\end{array}
\end{equation}
In contrast, for event horizons $C=0$ but away from equilibrium $\tl > 0$
(thanks to the second law) and so the horizon can still expand. 
For the timelike surfaces of the membrane paradigm none of $\tl$, $\tn$ or $C$ vanish. 
Equation \rf{AreaExp} will be very important in the following, also in cases
when $\tl$ is not exactly zero. 

Another gauge quantity 
\bea
\kappa_{\cV}  = - \cV^a n_b \nabla_a \ell^b
\eea
describes how the null normal evolves up $\triangle$. Under rescalings
\bea
\ell \rightarrow f(\lambda) \ell 
\Longrightarrow \; 
\kappa_\mathcal{V} \rightarrow f \kappa_\mathcal{V} + \frac{df}{d \lambda}  \, . \label{gaugekap}
\eea
 
In the case of an isolated horizon $\kappa_{\cV} = \kappa_{\ell}$, the familiar
\emph{surface gravity} that appears in the first law of black hole mechanics and
it is not hard to show that a zeroth law of black hole mechanics holds:
$\kappa_{\ell}$ is necessarily constant over the horizons \cite{IsoRef} (or section \ref{zeroth} of this paper),
though its exact value is fixed by the chosen normalization of the null vectors.

\subsubsection*{Constraints on the geometry}
These geometric quantities are not all independent. Instead they are linked by a set of constraint equations that ensure that the 
$S_\lambda$ fit smoothly together to form an $n$-tube and further that that tube must be embeddable in 
a larger $(n+1)$-dimensional spacetime (on which the Einstein equations hold). We now consider some of these constraints. 

First, the ``time'' rate of change of the two-metric on the $S_\lambda$ can be written in terms of the extrinsic curvature
\bea
\Lie_\cV \tq_{ab} = \tq_a^c \tq_b^d \Lie_\cV \tq_{cd} = 2 (k^{(\ell)}_{ab} - C k^{(n)}_{ab}) = (\tl - C \tn) \tq_{ab} 
+ 2 (\sigma^{(\ell)}_{ab} - C \sigma^{(n)}_{ab}) \label{deltaq}
\eea
and this is consistent with the already discussed (\ref{AreaExp}) if one takes the trace of both sides. 

More involved are the equations relating the derivatives of extrinsic curvature quantities to the rest of the geometry. Here we just 
list some of these results that will be used in our upcoming discussion of mechanics. Those interested in more details of the derivations
of these results can consult Appendix \ref{appA}.
First, for $\tl$
\bea
\Lie_{\cV} \tl &=& \kappa_{\cV} \tl - d^2 C + 2 \tom^a d_a C - C \left[\norm
  \tom \norm^2 - d_a \tom^a - \tilde{R}/2 + G_{ab} \ell^a n^b - \tl \tn \right]
\nn \\ 
& & - \left[ \frac{\tl^2}{n-1}   + \norm\sigma_{(\ell)}\norm^2 + G_{ab} \ell^a \ell^b \right] \label{explderiv}
\eea
and
\begin{eqnarray}
  {d}_{a}  \theta_{(\ell)} &=& \tl \tom_a
  2 ({d}_{b} - \tom_b) \sigma^{(\ell) b}_a
  - \frac{1}{n-1} \tilde{q}_{a}^{b}  G_{bc}\ell^{c} 
    - 2 \tq_a^b  \mathcal{C}_{bcde} \ell^{c} \ell^{d} n^{e} \, .
  \label{da_thetal} 
\end{eqnarray}
For $\tn$
\bea
  \Lie_\cV \theta_{(n)}  &=& -  \kappa_\cV \tn + 
    \left[\norm\tilde{\omega}\norm^2  + {d}_{a} \tilde{\omega}^{a} - \tilde{R}/2 
    + G_{ab} n^{a} \ell^{b} - \tl \tn \right] \nn
     \\
  && \quad 
  + C \left[  \frac{\theta_{(n)}^{2} }{n-1}+ \norm \sigma^{(n)}\norm^{2} 
    + G_{ab} n^{a} n^{b} 
     \right] \,   \label{expnderiv}
\eea
and
\begin{eqnarray}
  {d}_{a}  \theta_{(n)} &=& - \tn \tom_a
  2 ({d}_{b} + \tom_b) \sigma^{(n)  b}_a 
  - \frac{1}{n-1} \tilde{q}_{a}^{b} G_{bc} n^{c} 
    + 2 \tq_a^b \mathcal{C}_{bcde} n^{c} \ell^{d} n^{e}  \, \, . 
\end{eqnarray}
In these equations, $\norm \tom \norm^2 = \tom^a \tom_a$, $\norm \sigma^{(\ell,n)} \norm^2 = \sigma^{(\ell,n)}_{ab} \sigma_{(\ell,n)}^{ab}$, 
$\tilde{R}$ is the Ricci scalar on the $(n-1)$-surfaces, $G_{ab} = \mathcal{R}_{ab} - \frac{1}{2} \mathcal{R} g_{ab}$ is the Einstein tensor while $\mathcal{C}_{abcd}$ is the Weyl tensor. 
The only other equation that we will need is the ``time'' rate of change of $\tom_{a}$:
\bea
    \Lie_\cV \tilde{\omega}_{a}    &=&  {d}_a \kappa_\cV -  k^{(\ell)}_{ a b}
      \tilde{\omega}^b 
  - k^{(n)}_{ab}
  \left[{d}^b C - \tilde{\omega}^b C \label{dkappa}
    \right] \\
 & &   + \tilde{q}_{a}^{\; \; b} \left[ \frac{1}{(n-1)} G_{bc} (\ell^c + C n^c)     - \mathcal{C}_{bcde} {\cV}^{c}\ell^{d}n^{e} 
\right]    \nonumber \, .
 \end{eqnarray}
 
 As an immediate application of equation Eq.~(\ref{explderiv}), it can be used to (infinitesimally) quantify the lack of uniqueness of apparent
 horizons. Consider a particular marginally trapped surface $S_o$ and fix a scaling of $\ell^a$.  
Then one can try to solve (\ref{explderiv}) for $C$ to find a direction in which $S_o$ may be evolved while  maintaining $\tl =0$. 
If a solution exists then it is unique, but different scalings of $\ell^a$ will give rise to different $C$s and so different
possible directions of evolution;  in general, $S_o$ will be a member of an infinite number of different marginally trapped $n$-tubes. 
That said, (\ref{explderiv}) does strongly constrain the geometry of these possible evolutions from $S_o$. It can be shown that if one
chooses a particular $\triangle_o$, all other possible evolutions must lie partly in the causal past and partly in the causal future of $\triangle_o$.
Equivalently, all other $\triangle$ must ``interweave" with the original one (the original proof of this result may be found in \cite{GregAbhay}, but also see \cite{Booth:2006bn} for discussion in a language closer to that used here). 
 In particular if $\triangle_o$ is highly symmetric, then other
$\triangle$ cannot share that symmetry. Thus, in highly symmetric spacetimes it is usually possible to select a preferred marginally
trapped $n$-tube that shares those symmetries.

\subsection{Equilibrium states and the zeroth law}
\label{zeroth}

We now begin an examination of how the geometric calculations of the last section give rise to the laws of black hole mechanics. 
First consider equilibrium states and the zeroth law. 

Equilibrium states are characterized by null $n$-tubes on which $\tl = 0$ and $- G^a_{\phantom{a} b}
\ell^b$ is future-pointing and causal (a condition slightly weaker than the
dominant energy condition). These are a type of isolated horizons known as
\emph{non-expanding} horizons \cite{IsoRef} and as we now shall see include apparent
horizons that are in (possibly temporary) equilibrium with their surroundings as
well as event horizons in eternal equilibrium (such as the Kerr family of black
holes).

First if $\triangle$ is null, then $\mathcal{V}^a = \ell^a$ and $C=0$. Then  (\ref{explderiv}) reduces to the Raychaudhuri equation
\bea
\label{RaychaudhuriEqns}
\Lie_\ell \tl = \kappa_\cV \tl - \|\sigma^{(\ell)}\|^{2} - G_{ab} \ell^{a} \ell^{b} -  \theta_{(\ell)}^{2}/(n-1) \, , \label{Ray}
\eea
or with $\tl = 0$:
\bea
 \|\sigma^{(\ell)}\|^{2} + G_{ab} \ell^{a} \ell^{b} = 0 \, . 
\eea
On the geometry side it is clear that $\sigma^{(\ell)}_{ab} = 0$ and so the intrinsic geometry of 
$\triangle$ is time invariant
\bea
\Lie_{\ell} \tq_{ab} = \frac{1}{2} k^{(\ell)}_{ab} = \frac{1}{2} \tl \tq_{ab} + \sigma^{(\ell)}_{ab} = 0.  
\eea
Meanwhile, on the energy side, Raychaudhuri implies that $G_{ab} \ell^a \ell^b = 0$ and so by the energy condition 
$- G^a_{\phantom{a} b} \ell^b = \mu \ell^a$ for some function $\mu$: there are no matter flows across the horizon. 

If one further assumes that  the $\tom_a$ and $\kappa_\ell$ components of the extrinsic geometry are time-invariant
($\Lie_\ell \tom_a = 0$ and $\Lie_\ell \kappa_\ell = 0$) we have a \emph{weakly isolated horizon}. Then, by (\ref{da_thetal})
$\tq_a^b  \mathcal{C}_{bcde} \ell^{c} \ell^{d} n^{e} = 0$ and so gathering many of the previous results together, 
(\ref{dkappa}) implies that $d_a \kappa_\ell = 0$ and so the surface gravity is constant over the horizon. This is the zeroth law
of black hole mechanics. 

There is also a phase space version of the first law of black hole mechanics which examines how physical quantities change across
the phase space of isolated horizons \cite{IsoRef}. It follows from a careful Hamiltonian analysis of isolated horizons however it is 
beyond the scope of this paper. Here we will instead focus on dynamic versions of the first and second law, starting with event horizons. 

\subsection{The first and second law for  event horizon mechanics}

Next consider dynamic event horizons -- null $n$-tubes with $\tl \neq 0$ which satisfy appropriate teleological boundary 
conditions. Then the Raychaudhuri equation (\ref{Ray}) multiplied by the area element can be rewritten as 
\bea
\left(\kappa_{\ell} +  \left[ \frac{n-2}{n-1} \right] \tl \right) \Lie_\ell \vS = \Lie_{\ell} \left( \Lie_{\ell} \vS  \right) 
 + \vS \left(  \norm \sigma^{(\ell)} \norm^2 + G_{ab} \ell^a \ell^b \right ).   \label{RayRewrite}
\eea
Forms of the first law of black hole mechanics may be derived from this in two distinct ways. 
First, the \emph{teleological first law}.  On integrating (\ref{RayRewrite}) over the horizon between the slices $v=v_1$ and $v=v_2$, 
it becomes
\bea
\int_{v_1}^{v_2}   \mspace{-12mu} d\lambda \int d^2 x  \left(\kappa_{\ell} +   \left[ \frac{n-2}{n-1} \right] \tl \right) \Lie_\ell \sqrt{\tq} 
 = \left. \frac{da}{dv} \right|^{S_2}_{S_1}
+ \int_{v_1}^{v_2}  \mspace{-12mu} d\lambda \int \vS \left(  \norm \sigma^{(\ell)} \norm^2 + G_{ab} \ell^a \ell^b \right ),
\eea
where $a$ is the area of the horizon. In the case where the event horizon transitions between (near) equilibrium states on which 
$\dot{a} \approx 0$ and does so slowly (so that $\tl$ is small relative to $\kappa_\ell$) this gives the integrated
first law
\bea
\int_{v_1}^{v_2}  \mspace{-12mu} d\lambda \int d^2 x  \left( \kappa_{\ell}  \Lie_\ell \sqrt{\tq} \right)
\approx
 \int_{v_1}^{v_2}  \mspace{-12mu} d\lambda \int \vS \left(  \norm \sigma^{(\ell)} \norm^2 + G_{ab} \ell^a \ell^b \right ).
\label{EHev}
\eea
As we saw in section \ref{HorizonIntro}, contrary to initial intution, influxes
of matter or gravitational radiation do not drive event horizons expansions, but
instead curtail existing expansions (this can also be seen in the original
Raychaudhuri equation (\ref{Ray}) where an increasing matter  
or shear flux clearly results in a decrease in $\Lie_\ell  \tl$).  In this
version of the first law, the integration erases this feature  
and leaves behind what appears to be a more intuitive first law (first derived in \cite{hawkhartle}). 

That said, there is a regime where a more intuitive version of the first law holds. This is the second version of the first law which we will
refer to as the \emph{causal first law} since in this case fluxes appear to drive the expansion.  If
\bea
\Lie_\ell (\Lie_\ell  \sqrt{\tq}) \ll \Lie_\ell \sqrt{\tq} \ll \sqrt{\tq} \kappa_\ell \;  \Leftrightarrow \; 
\Lie_\ell \tl \ll \tl \ll \kappa_\ell  \, , \label{DerAssump}
\eea
then (\ref{Ray}) reduces to:
\bea 
\kappa_{\ell} \Lie_\ell \vS \approx \vS \left(  \norm \sigma^{(\ell)} \norm^2 + G_{ab} \ell^a \ell^b \right ) \, , \label{pbp_1st}
\eea
and a point-by-point version of the first law (approximately) holds where the fluxes (appear to) drive the expansion. Note that 
(\ref{DerAssump}) assumes that area rates of change are small and the second
derivative of area is much smaller than the first. So in cases  
where a derivative expansion can be made in the area element (which will be the
case in section \ref{EH}), this local version of the first 
law holds for event horizons.

It is important to emphasize that both of these versions of the first law apply to event horizons that are evolving slowly. In the first
case we assumed the expansion $\tl \ll \kappa_{(\ell)}$ while in the second we assumed the even more restrictive (\ref{DerAssump}). 
In this sense, both laws apply to horizons that are close to equilibrium. 
This should not come as a surprise. In regular thermodynamics the $dE = TdS$ form of the first law (as opposed to a 
more general statement about conservation of energy) applies only in quasi-equilibrium situations. As such it is natural to expect the 
same thing for black hole mechanics. This will be a recurring idea in the rest of this paper. 

Finally, the Raychaudhuri equation also lies at the root of the second law. For any congruence of null geodesics we can alway find an 
affine scaling of the null vectors so that $\kappa_{\ell} =  0$. Then, if $\tl <
0$ and the null energy condition holds, (\ref{Ray}) implies that 
\bea
\Lie_\ell \tl \leq - (1/2) \theta_{(\ell)}^{2}
\eea
and so a congruence with $\tl < 0$ necessarily includes caustics. 
This is the origin of the second law of event horizon mechanics. Other results 
show that event horizons don't have caustics and therefore by the preceding argument they must
have $\tl \geq 0 $ everywhere.  In turn this implies that  $\Lie_\ell \sqrt{\tq} = \sqrt{\tq} \tl \geq 0$; event horizons have
non-decreasing area.

\subsection{The first and second law of apparent horizon/FOTH  mechanics}

We now turn to the physics of dynamical $n$-tubes which are foliated by $\tl = 0$ surfaces. In contrast to the last section, 
we begin with the second law. 

Just as the Raychaudhuri equation (\ref{Ray}) was key to event horizon dynamics its generalization Eq.(\ref{explderiv}) is the 
keystone equation for dynamical apparent horizons. Setting $\tl = 0$ we get a partial differential equation giving allowed values of
$C$ and if one also imposes $\tn < 0$ and 
\bea
\Lie_n \tl = \norm \tom \norm^2 - d_a \tom^a - \tilde{R}/2 + G_{ab} \ell^a n^b  < 0 \label{LntL}
\eea
 as motivated in section \ref{HorizonIntro} many results follow.  
To begin, the signature of $\triangle$ is strongly constrained:
\begin{enumerate}
\item $\triangle$ is null (and so isolated) if and only if no matter or gravitational wave flux crosses the horizon: $C= 0$ $\Leftrightarrow$ $T_{ab} \ell^a \ell^b = 0$
and $\sigma^{(\ell)}_{ab} = 0$. 
\item $\triangle$ is timelike (and so decreasing in area) only if the energy conditions are violated: $C < 0$ $\Rightarrow$ $T_{ab} \ell^a \ell^b < 0 $. This can happen, for example, in the presence of Hawking radiation. 
\item Under all other circumstance, interactions with its environment cause $\triangle$ to be spacelike 
(and so expanding in area). In this case, the horizon 
is said to be dynamical \cite{DynRef} and this is the \emph{second law of dynamical horizon mechanics}. 
\end{enumerate}
This set of results was first shown in \cite{hayward} but has since been reproved many times. In the special case of $C$
constant over each $S_\lambda$ this can be quite easily be seen from (\ref{explderiv}) with the understanding that (\ref{LntL}) holds.
More generally one applies a maximum principle to show these results (see, for example, \cite{Booth:2006bn}).

Next, the first law. As for event horizon mechanics, we expect this to hold close to equilibrium and so the 
first task is to define ``near'' equilibrium for FOTHs.  
Though the intuition is fairly clear, the implementation is not so
straightforward. A near equilibrium FOTH should be:  
1) slowly expanding and 2) almost null. Neither of these is trivial to
characterize. With respect to the first condition, the second law says that   
dynamical FOTHs are spacelike and so there is no natural flow of time along
the horizon against which one can judge the rate 
of expansion of the horizon. With respect to the second condition, spacelike and
null normal vectors are qualitatively different with 
squared norms that are respectively positive and zero. While one might naively
say that an approach to isolation could be tracked 
by following how the norm of the normal goes to zero, this runs into
difficulties.  The only natural scaling of a spacelike vector is  
the unit-scaling which manifestly will not approach zero but instead become
undefined if the normal becomes null.  
Thus one needs some extra structure in order to specify when a spacelike vector 
is ``almost'' null -- either an alternative scaling or some new idea. 

Fortunately a formalism already exists that deals with these problems:
\emph{slowly evolving horizons}
\cite{Booth:2003ji,Kavanagh:2006qe,Booth:2006bn}. This was
developed with the twin goals of getting a better understanding of
quasi-equilibrium black hole mechanics and developing a set of tools to track the
approach (or departure) of a horizon from equilibrium. Though originally
intended for compact horizons in regular four-dimensional spacetimes it is
easily adapted to (non-compact) black branes in arbitrary number of dimensions.

As noted in \rf{AreaExp}, with $\tl=0$ the rate of change of horizon area is related to the expansion of the ingoing null normal by 
\bea
\Lie_{\cV} \sqrt{\tilde{q}} = - C \tn \sqrt{\tq} \, . 
\eea
The definition of $\cV^a$ \rf{cV} implies that if one relabels the foliation
surfaces with $\lambda \rightarrow \tilde{\lambda}$ then  
$\ell^a \rightarrow f \ell^a$, $n^a \rightarrow n^a/f$ and $C \rightarrow f^2 C$
where $f = d \tilde{\lambda}/ d \lambda$. As such, the  
scaling dependence of the expansion may be isolated by writing
\bea
\Lie_{\cV} \sqrt{\tilde{q}} =  \sqrt{\tq} || \mathcal{V} || \left(-
\sqrt{\frac{C}{2}} \tn  \right)   \equiv  \sqrt{\tq}  || \mathcal{V} ||  
\theta_{(\widehat{\cV})}
\, , 
\eea
with the scaling dependence restricted to $|| \cV || = \sqrt{2C}$. The scaling-independent part is the expansion 
$\theta_{(\widehat{\cV})}$ associated with the unit normalized evolution vector $\widehat{\cV}^a = \mathcal{V}^a / || \cV ||$. 
Note that even though $\widehat{\cV}^a$ is not defined in the null limit, $\theta_{(\widehat{\cV})}$ is well-defined and has
the desired limiting behaviour:  $\theta_{(\widehat{\cV})} \rightarrow 0$ as $\mathcal{V}^a$ becomes null \cite{Booth:2007Short}.

Thus there is a reasonable way to characterize slow expansion. To strengthen the ``almost-null'' analogy
one should also require that, to first order, the evolution of the geometry of the foliations of the horizon be
characterized by the expansion and shear associated with $\ell^a$ (as they would
be for a truly null surface). Similarly the flux should be approximately that of a null surface. Thus one would like to find the  
conditions under which 
\begin{eqnarray}\label{Ltq_sig}
  &\Lie_{\cV} \tq_{ab} &=   \left(\tl  -  C \tn \right) \tq_{ab}  + 2  (\sigma^{(\ell)}_{ab} -  C\sigma^{(n)}_{ab}  ) 
  \approx  2 \sigma^{(\ell)}_{ab} \,  ,  
  \mbox{and} \\
&T_{ab} \cV^a  \tau^b &= T_{ab} (\ell^a - C n^a)(\ell^b + C n^b) \approx  T_{ab} \ell^a \ell^b 
\, , \label{MatFlux}
\eea
(since $\tl$ vanishes). Further, one would expect these quantities to be small since on a truly isolated horizon both shears  vanish. 

Following \cite{Booth:2006bn} this can all be put on a better mathematical footing by making a few assumptions about the various 
quantities appearing in Eq. (\ref{Ltq_sig}). Using the notation $A \lesssim B$ to indicate that $A \leq k_o B$ for some $k_o$ of order one, 
one assumes that all quantities are bounded relative to some length scale $\mathscr{L}$:
\bea
| \tilde{R} | \, , \,  \tom^a \tom_a \, , \,  |d_a \tom^a| 
 \; \mbox{and} \;  |T_{ab} \ell^a n^b| 
\lesssim \frac{1} {\mathscr{L}} \, , 
\eea
and similarly derivatives of $C$ are commensurate
\bea
\norm d_a C \norm \lesssim \frac{C_{max}}{\mathscr{L}} \; \; \mbox{and} \; \; 
\norm d_a d_b C \norm \lesssim \frac{C_{max}}{\mathscr{L}^2}  \, . 
\eea
These are essentially assumptions that all quantities are of a ``reasonable'' size and the geometry not be too extreme. 
They are actually quite weak and, for example, in four-dimensions all members of the
Kerr family of solutions satisfy the first condition  
if $\mathcal{L}$ is taken as the areal radius. 

Then, the rate of expansion is characterized by 
an \emph{evolution parameter} $\epsilon$ defined by:
\begin{equation}  \epsilon^2 /  \mathscr{L}^2  =  \mbox{Maximum} 
  \left[C \left( \norm\sigma^{(n)}\norm^2 
  + T_{ab} n^a n^b 
  + \tn^2/2 \right) \right] \label{epsilon}
  \, .
\end{equation}
Note that if one wishes to allow for energy condition violating matter (so that $C< 0$ is possible) then it is necessary to take the 
absolute value of the right-hand side to ensure that $\epsilon^2 > 0$. 

This quantity can be thought of as a generalization of $\theta_{\hat{\cV}}^2$ and is independent of the scaling of the null vectors. 
If $\epsilon \ll 1$,  Eq.~(\ref{explderiv}) can be used to show that 
\bea
\norm \sigma^{(\ell)}_{ab} \norm \lesssim \frac{\sqrt{C_{max}}}{\mathscr{L}} \; \; \mbox{while}
\; \; C \norm\sigma^{(n)}_{ab} \norm \, \, \mbox{and} \, \,  C |\tn|  \lesssim \epsilon \frac{\sqrt{C_{max}}}{\mathscr{L}} \,  
, 
\eea
and so Eq.~(\ref{Ltq_sig}) is quantified:
\bea
\Lie_{\cV} \tq_{ab} =\underbrace{\sigma^{(\ell)}_{ab}}_{O(\sqrt{C}) } + \underbrace{\left( - C \tn \tq_
{ab} - C\sigma^{(n)}_{ab}\right)}_{ O(\epsilon \sqrt{C})} \, . \label{LcVex} 
\eea

These ideas are consolidated to get the first part of the definition of a near-equilibrium FOTH:

\smallskip
\noindent
{\bf Definition:} Let $\triangle H$ be a section of a $n$-dimensional FOTH
foliated by (non-compact) spacelike $(n-1)$-surfaces $S_\lambda$ so that 
$\triangle H = \{ \cup_v S_\lambda : \lambda_1 \leq \lambda \leq \lambda_2\}$. Further let
$\cV^a$ be an evolution vector field that generates the foliation so
that $\Lie_{\cV} \lambda = \alpha(\lambda)$ for some positive function $\alpha(\lambda)$, and
scale the null vectors so that $\cV^a = \ell^a - C n^a$. Then $\triangle
H$ is a \emph{slowly expanding horizon} relative to the length scale $\mathscr{L}$ 
if 

\begin{enumerate}[(i)]
\item the \emph{evolution parameter} $\epsilon \ll 1$ where 
\begin{equation}
  \epsilon^2 /  \mathscr{L}^2  =  \mbox{Maximum} 
  \left[C \left( \norm\sigma^{(n)}\norm^2 
  + T_{ab} n^a n^b 
  + \tn^2/2 \right) \right] 
  \, ,
\end{equation}
\item $\displaystyle |\tilde{R}|$ and $\norm \tom \norm^{2}  \lesssim 1/\mathscr{L}^2 $  
and $T_{ab} \ell^a n^b \lesssim 1/\mathscr{L}^2$
\item $(n-1)$-surface derivatives of horizon fields are at most of the same
order in $\epsilon$ as the (maximum of the) original fields.  For
example, $\displaystyle \norm d_a C\norm \lesssim  {C_{max}}/{\mathscr{L}}$, 
where
$C_{max}$ is largest absolute value attained by $C$ on $S_\lambda$. 
\end{enumerate}

Compared to the original definition (for three-dimensional horizons in
four-dimensional spacetimes \cite{Booth:2003ji,Booth:2006bn}), the only changes
are in the dimension and switching from compact to non-compact horizons. A
secondary change following from these is that the area of the (compact)
two-dimensional horizon cross-sections was used in the original definition to
set the scale $\mathscr{L}$. For non-compact horizons this is no longer feasible
and so the length scale must be set differently. For the black branes of this paper, the
only scale is set by the cosmological constant and so we define  $\Lambda
= - (n-1) n / 2 / \mathscr{L}^{2}$ (the relevant scale is the AdS radius);
for the coordinates that we are using is equivalent to setting $\mathscr{L}=1$. 

To move from geometry to mechanics, the formalism requires that there exist
a scaling of the null vectors that satisfies the following  
(again slightly modified) conditions:

\smallskip
\noindent
{\bf Definition:} 
\noindent A slowly expanding horizon is said to be \emph{slowly evolving}
if there exists a scaling of the null vectors such that $C \lesssim
\epsilon^2$ and: 
\begin{enumerate}[(i)]
\item $\displaystyle \norm\Lie_{\cV} \tom_a \norm$ and $\displaystyle |\Lie_{\cV}
\kappa_{\cV}|  \lesssim  {\epsilon}/{\mathscr{L}^2}$ and 
\item $\displaystyle |\Lie_{\cV} \tn| \lesssim {\epsilon}/{\mathscr{L}^2}$. 
\end{enumerate}

Scaling the null vectors so that $C \lesssim \epsilon^2$ is motivated
by a couple of considerations. First, it means that the norm $|| \cV || = 2 C \lesssim
\epsilon^2$ and so in an asymptotic approach to isolation the tangent
vector becomes null in an orderly fashion (for example it doesn't diverge
in the limit). Second with this scaling, $\Lie_{\cV}$ ``time''-derivatives will properly 
reflect the slowly evolving nature of the horizon; for example 
$\Lie_{\cV} \sqrt{\tq} = -C \theta_{(n)} \sqrt{\tq} \lesssim \epsilon^2/ \mathcal{L}$ and
so the area expansion is also slow. 

The other conditions are motivated by the isolated horizon formalism. For a
horizon in equilibrium with its surroundings, foliations can always be found so
that both of these vanish and then, as we have seen, the zeroth law of isolated
horizon mechanics directly follows. Similarly for slowly evolving,
near-equilibrium horizons, these conditions are sufficient to enforce an
approximate zeroth law: surface gravity is constant across slices to order
$\epsilon$. The derivation is essentially the same as that for the true first law -- one 
simply applies the assumptions to (\ref{dkappa}) this time setting quantities to be small
rather than zero. Note though that in cases of high symmetry (such as those that will be considered later
in this paper), it often turns out that those symmetries will force $d_a \kappa_\cV = 0 $ exactly, independently of
these considerations. 

Finally one can combine (\ref{explderiv}) and (\ref{expnderiv}) to show that 
\bea
\kappa_{\cV} \theta_{\cV} &=& \Lie_\cV \tl + C \Lie_{\cV} \tn 
      + d_a (d^a C - 2C \tom^a) \label{kapT} \\
           & & 
   + \sigma^{(\tau)} \mspace{-6mu} : \mspace{-4mu} \sigma^{(\cV)} 
   + G_{ab} \cV^a \tau^b 
   + \theta_{(\cV)} \theta_{(\tau)}/(n-1) \, , 
\nn
\eea
where $\tau^a = \ell^a + C n^a$. Now for a slowly evolving horizon $\tl$ and
$\Lie_\cV \tl$ vanish 
and on including the other assumptions and approximations this equation
reduces to 
\bea
\kappa_o \Lie_{\cV} \sqrt{\tilde{q}} \approx \sqrt{\tilde{q}} \left( \norm \sigma^{(\ell)} \norm^2 + T_{ab} \ell^a \ell^b \right) \, , \label{FirstLaw}
\eea
where $\kappa_o$ is the lowest order of the surface gravity expansion on the
slice: $\kappa_{\cV} \approx \kappa_o + \epsilon \kappa_1$.
This is the first law of slowly evolving FOTHs/dynamical horizons.

\subsection{The first and second law of alternate ``horizon" mechanics}
\label{AlternateHOR}

In the previous subsections we have seen that both apparent and event horizons give rise to reasonable notions 
of black hole mechanics with the same zeroth law, similar first laws, and different second laws (in that the horizons
have different surface areas and so different notions of entropy). Now compared to isolated horizons and stationary black holes, 
event horizons drop the $\tl = 0$ requirement while dynamical apparent horizons are no longer null. However both return (at
least asymptotically) to become isolated horizons in the equilibrium limit. With two candidates in hand, it is then natural to
consider other $n$-tubes which might act as horizons: correctly interpolating between equilibrium states and obeying the laws
of mechanics. Guided by the example of slowly evolving horizons we will continue to consider the near-equilbrium limit: essentially 
we will consider slowly evolving almost-horizons for which $\tl \approx 0$. The guiding principles in defining slowly evolving almost-horizons will be: 
\begin{enumerate}[{P}1.]
\item The ``horizon'' $n$-tube should be a one-way membrane in the sense that no causal signal from a trapped surface should be 
able to cross it in the direction of infinity. \label{oneway}
\item Under equilibrium conditions the $n$-tube should match or at least asymptote to an isolated horizon. That is, the usual notion
of equilibrium entropy should be recovered. \label{asympt}
\item Near equilibrium the $n$-tube should be ``almost'' a slowly evolving horizon (and so ``almost'' isolated). \label{SEH}
\item Near equilibrium the surface gravity $\kappa_{\cV}$ should be almost constant. \label{0Law}
\item Near equilibrium there should be a first law of the form (\ref{FirstLaw}). \label{1Law}
\item The area should be non-decreasing. \label{2Law}
\item In the apparent and event horizon limits, the conditions should reduce the known laws for those cases. 
\end{enumerate}
Guided by the experience from slowly evolving horizons, we now propose a class
of $n$-tubes that satisfy these conditions. As before we will need to be careful
to make sure that everything remains independent of the scaling of the null
normals but now have the extra challenge  
when setting conditions that we may have $\tl \neq 0$ and $C = 0$ simultaneously. Thus, unlike for slowly evolving horizons we can't
use powers of $C$ to remove scaling invariance. Instead we will use $\tn$ which
for standard black hole/brane solutions is non-zero and  
of finite size when $\tl = 0$. 

The following structures will meet all of the guiding principles. 

\smallskip
\noindent
{\bf Definition:}
Let $\triangle H$ be a section of a $n$-tube
foliated by spacelike $(n-1)$-surfaces $S_\lambda$ so that
$\triangle H = \{ \cup_v S_\lambda : \lambda_1 \leq \lambda \leq \lambda_2\}$. Further let
$\cV^a$ be an evolution vector field that generates the foliation so
that $\Lie_{\cV} \lambda = \alpha(\lambda)$ for some positive function $\alpha(\lambda)$, and
scale the null vectors so that $\cV^a = \ell^a - C n^a$. 
Then $\triangle H$ is a \emph{near-equilibrium $n$-tube (NENT)} relative to the length scale $\mathscr{L}$ if: 
\begin{enumerate}[(i)]
\item the \emph{evolution parameter} $\epsilon \ll 1$ where 
\begin{equation}
  \epsilon^2 /  \mathscr{L}^2  =  \mbox{Maximum} 
  \left[|\tn \tl | + | C | \left(\tn^2 +  \norm \sigma^{(n)} \norm^2 + T_{ab} n^a n^b \right) 
  \right] 
  \, , \nn
\end{equation} \label{evPar}
\item $\displaystyle |\tilde{R}|$ and $\norm \tom \norm^{2}  \lesssim 1/\mathscr{L}^2 $  
and $T_{ab} \ell^a n^b \lesssim 1/\mathscr{L}^2$
\item $(n-1)$-surface derivatives of horizon fields are at most of the same
order in $\epsilon$ as the (maximum of the) original fields.  For
example, $\displaystyle \norm d_a C\norm \lesssim  {C_{max}}/{\mathscr{L}}$, 
where
$C_{max}$ is largest absolute value attained by $C$ on $S_\lambda$. 
\end{enumerate}
Further the scaling of the null normals may be chosen so that 
\begin{enumerate}[(i)]
\addtocounter{enumi}{3}
\item $\kappa_{(\cV)}, |\tn| \sim  1/\mathscr{L}$, and $\Lie_n \tl < 0$, $\tl - C \tn > 0$ \label{HorCons}
\item $|\Lie_\cV \tl| \lesssim \epsilon^3/\mathscr{L}^3$
\item $\displaystyle \norm\Lie_{\cV} \tom_a \norm$ and $\displaystyle |\Lie_{\cV}
\kappa_{\cV}|  \lesssim  {\epsilon}/{\mathscr{L}^2}$ and 
\item $\displaystyle |\Lie_{\cV} \tn| \lesssim {\epsilon}/{\mathscr{L}^2}$. 
\end{enumerate}

Let us consider this definition and how it meets the guiding principles. First if $\tl = 0$ on $\triangle H$, then 
it reduces to the definition of a slowly evolving horizon. Similarly if $C = 0$ it will reduce to the special case of an
event horizon for which the point-by-point first law (\ref{pbp_1st}) holds. Further if $\epsilon = 0$ then $\tl=0$ and $C=0$ 
and we are back to an isolated horizon. 

We now consider where this definition differs from that of a slowly evolving horizon and the resulting implications. 
First (\ref{evPar}) contains an extra term and this has the joint purpose of ensuring that the outward null-expansion is small and 
also that $\epsilon$ is defined even for null horizons where $C=0$. Further, applying it along with conditions (ii)-(v) to 
the expression (\ref{explderiv}) for $\Lie_\cV \tl$  one can show that 
\bea
\norm \sigma^{(\ell)} \norm^2 \lesssim \epsilon^2/\mathscr{L}^2 \; \; \mbox{and} \; \;  T_{ab} \ell^a \ell^b \lesssim \epsilon^2/\mathscr{L}^2
\eea
in essentially the same way that these quantities were bounded for isolated and slowly evolving horizons. Note, too that by bounding 
$|\tl \tn|$ to be of the order of $\epsilon^2$ we retain the result that to order $\epsilon$ the horizon will evolve as a $\tl = 0$ null surface:
\bea
\Lie_{\cV} \tq_{ab} \approx  \sigma^{(\ell)}_{ab} \, . 
\eea
If $|\tl \tn| $ was of order $\epsilon$ then this relation would be lost and a host of difficulties would follow, for example in deriving the first law.
Similarly, to order $\epsilon^2$ the matter flux across the horizon will also be that for a 
null surface:
\bea
T_{ab} \cV^a \tau^b \approx T_{ab} \ell^a \ell^b \, .  
\eea

The scaling condition $\tn \sim -1/\mathscr{L}$ is essentially that used in the standard coordinate descriptions of exact black hole solutions
and also for spacelike or timelike horizons implies $C \sim \epsilon^2$. The second law is put in by hand in the assumption 
$\tl - C \tn > 0$ however note that if $\tl > 0$ (outside a FOTH) then timelike $\triangle H$ with $C< 0$ are allowed. 
At the same time (vi-vii) guarantee an approximate zeroth law by essentially the same arguments that we saw for isolated and slowly 
evolving horizons. Similarly, one may rerun the slowly evolving arguments by applying our conditions to (\ref{kapT}) to 
rederive the first law (\ref{FirstLaw}) for these ``horizons''. 

The only condition left to be considered is P1. This is clearly okay if $\triangle H$ is spacelike or null. Intuitively however it should also be
okay if $C < 0$. As discussed in the previous paragraph, this is only possible if $\tl > 0$ -- outside any FOTH. So, signals could
pass out of the NENT -- but only from a boundary layer outside the trapped region. This is similar to the situation in the membrane 
paradigm where the boundary is taken to be a timelike surface.

\section{Gravity dual to Bjorken flow}
\label{gradubi}

\subsection{Fluid/gravity duality and ``alternate horizons''}

One of the most important recent insights within the AdS/CFT correspondence is a
gravity dual formulation of (conformal) relativistic hydrodynamics
\cite{Bhattacharyya:2008jc}. It has been known for many years that black holes
are thermal in nature. In the context of gauge/gravity duality thermal states
of certain strongly coupled quantum field theories have been understood in terms
of asymptotically AdS black brane geometries. From the perspective of
fluid/gravity duality hydrodynamics 
is thought of as long wave-length dynamics of
non-equilibrium branes. Dissipative effects in hydrodynamics imply entropy
production. On the other hand, entropy on the gravity side had been linked for
many years with event horizons and area increase theorems with the second law of
thermodynamics. This leads to a unique\footnote{Up to horizon-boundary mapping} 
identification of the entropy also in the context of the gravity picture of
hydrodynamics\cite{Bhattacharyya:2008xc}, in contrast with the
phenomenological definition of the boundary entropy where some freedom remains
\cite{Bhattacharyya:2008xc,Romatschke:2009kr}. However, as reviewed in the last section, there are many
notions of horizon on the gravity side. These lead to distinct definitions of
entropy which coincide in the equilibrium situation, which makes it interesting to
first investigate the conditions under which a gravitational system is close to
equilibrium. The answer can be given with the help of the slowly-evolving
formalism generalized to the AdS case in the previous section. After identification of the
near-equilibrium situations one may proceed to understand various notions of horizons
and identify the freedom of definition of entropy in the gravity dual to
hydrodynamics. Since the entropy is an infrared concept, its local
identification on the gravity side relies on considerations of surfaces close to
the horizon (or, in other words, in the region corresponding to the thermal scale
on the gauge theory side). The  
event horizon itself is a teleological concept, and as such is sometimes hard to
deal with. The surfaces which capture the notion of entropy are better
characterized by $\tl \approx 0$ (the ``alternate horizons'' from the
previous section). This uses the fact that close to 
equilibrium (i.e. in the case of slow evolution) apparent and event horizon
should be close to each 
other.

Understanding this in the general case might be shadowed by significant
technical details. However, the fluid/gravity duality links solutions of
relativistic conformal hydrodynamics with particular solutions of Einstein
equations with negative cosmological constant. One such solution is
boost-invariant expansion known as boost-invariant flow \cite{Bjorken}. It is
complicated enough to capture many non-trivial features of the hydrodynamics and
at the same time simple enough to enable efficient treatment of the
5-dimensional geometry in the horizon region. As such it serves as the a
laboratory for exploring different notions of entropy and the corresponding
notions in the hydrodynamics of 4-dimensional supersymmetric Yang-Mills plasma.

\subsection{Bjorken flow}

Boost-invariant flow is an one-dimensional expansion of plasma with the
boost symmetry along the expansion axis. If $x^{0}$ denotes lab frame time
and $x^{1}$ is the coordinate along the expansion axis, then transformation
to more convenient coordinates -- proper time $\tau$ and rapidity $y$ --
takes the form
\begin{eqnarray}
x^{0} &=& \tau \cosh{y} \mathrm{,}\nonumber \\
x^{1} &=& \tau \sinh{y} \mathrm{.}
\end{eqnarray}
The assumption of boost-invariance is based on Bjorken's observation that
multiplicity spectra does not depend on rapidity in the mid-rapidity
region. The most general traceless energy-momentum tensor, which is
boost-invariant can be expressed solely in terms of the energy density
$\epsilon \left(\tau \right)$
\begin{equation}
\label{BIstresstensor}
T_{\mu \nu} = \mathrm{diag}\left\{\epsilon\left(\tau\right), \, - \tau^{3}
\epsilon'\left( \tau \right) - \tau^{2} \epsilon \left(\tau \right), \,
\epsilon\left( \tau \right) + \frac{1}{2} \tau \epsilon'\left(\tau\right),
\, \epsilon\left( \tau \right) + \frac{1}{2} \tau
\epsilon'\left(\tau\right)\right\}
\end{equation}
which is a function of proper time only (in boost-invariant setup no
physical quantity can depend on rapidity) \cite{Janik:2005zt}. 

Applicability of the hydrodynamic description
means that the system is fully described by the four-velocity $u^{\mu}$
($u_{\mu} u^{\mu} = -1$) and temperature $T\left(\tau\right)$. Boost-invariance
implies that $u^{\mu} = \left[ \partial_{\tau}\right]^{\mu}$ and, as expected on
the basis of the form of the energy momentum tensor, the full dynamics  
is encoded in the dependence of the energy density on $\tau$. Since for a
conformal fluid the
equation of state is $p=1/3 \epsilon$, one has
\bea
\label{edens}
\epsilon &=& e_0 T^{4} \\
\label{entrodens}
s &=& \frac{4}{3} e_0 T^3 \ , 
\eea
where $s$ is the usual thermodynamic entropy density.

The general form of second order 
hydrodynamics equations (see \cite{Baier:2007ix}) is 
\begin{eqnarray}
\partial_{\tau} \epsilon &=& - \frac{4}{3} \frac{\epsilon}{\tau} +
\frac{\Phi}{\tau} \mathrm{,} \\ \nonumber \tau_{\Pi} \partial_{\tau} \Phi
&=& \frac{4}{3} \frac{\eta}{\tau} - \Phi - \frac{4}{3}
\frac{\tau_{\Pi}}{\tau} \Phi - \frac{1}{2} \frac{\lambda_{1}}{\eta^{2}}
\Phi^{2} \mathrm{.}
\end{eqnarray}
Here $\Phi$ is yy component of the shear tensor $\Pi^{\mu \nu}$. These equations
determine the proper-time evolution of 
the energy density, which in turn determines the dependence of temperature of $\tau$. 

Substituting the form of the energy density 
\rf{edens} (together with the
equation of state) into these equations 
leads to the following solution for the temperature as a function of
proper-time:  
\begin{equation}
\label{hydrotemp}
T\left( \tau \right) = \frac{\Lambda}{\tau^{1/3}} \left\{1 -
\frac{1}{\Lambda \, \tau^{2/3}} \cdot \frac{\eta_{0}}{\sqrt{2} \, 3^{1/4}
  \pi}+ \frac{1}{\Lambda^{2} \, \tau^{4/3}} \left(
\frac{\lambda^{\left(0\right)}_{1}}{3 \sqrt{3} \, \pi^{2}} - \frac{\eta_{0}
  \tau^{\left(0\right)}_{\Pi}}{3 \sqrt{3} \, \pi^{2}} \right)+ \ldots
\right\}
\end{equation}
where $\Lambda$ is a scale fixed by the initial conditions and the only
arbitrary number in the construction\footnote{It is easy to understand the
  presence of $\Lambda$ if one considers the dilatation $x^{\mu}
  \rightarrow \alpha \, x^{\mu}$. Then $\tau \rightarrow \alpha \, \tau$
  and $\Lambda \rightarrow \alpha^{-2/3}$ which leads to $\epsilon\left(
  \alpha \tau \right) = \alpha^{-4} \epsilon\left( \tau \right)$ which is
  the correct scaling}. In much of the literature (such as
\cite{Janik:2005zt,Janik:2006ft,Heller:2007qt,Benincasa:2007tp}) the choice
$\Lambda = \frac{\sqrt{2}}{3^{1/4} \pi}$ is made. The constants $\eta_{0}$,
$\tau^{\left(0\right)}_{\Pi}$, $\lambda^{\left(0\right)}_{1}$ are various
transport coefficients from the 
first ($\eta_{0}$) and second order viscous hydrodynamics
\footnote{More precisely these are dimensionless numbers related to the transport coefficients by the relations
\protect \begin{eqnarray}
\eta &=& \eta_{0} e_{0} T^{3} \mathrm{,}\nonumber \\ 
\tau_{\Pi} &=& \tau_{\Pi}^{0} T^{-1} \mathrm{,}\nonumber \\
\lambda_{1} &=& \lambda_{1}^{0} e_{0} T^{2} \mathrm{,} 
\end{eqnarray} where $e0$ is defined as previously by $\epsilon = e_{0} T^{4}$}.
They are
universal numbers related to the microscopic physics of the underlying
quantum field theory. Their presence signals the dissipative nature of the
flow -- the entropy production.

\subsection{Validity of the hydrodynamic description}

The modern view of hydrodynamics is similar to that of effective field
theory. It is a phenomenological description of phenomena on scales much larger
than those of their microscopic dynamics, constructed as a systematic expansion
in gradients. In the context of boost-invariant flow this translates into an
expansion in powers of $1/\tau^{2/3}$. As in the case of any perturbative
expansion, one needs to observe the regime where the expansion can be reasonably
expected to apply. A criterion for this is that the subleading terms in the
expansion be smaller than the leading order. In the context of Bjorken flow this
can be understood as a condition on the minimal time when the expansion can be
trusted. For the first subleading term in \rf{hydrotemp} to be smaller than the
leading order\footnote{The temperature has been chosen here because it 
enters the definition of the gradient expansion.} one needs $\tau > \tau_{min}$,
where 
\bel{taumin}
\frac{1}{\Lambda \, \tau_{min}^{2/3}} \cdot \frac{\eta_{0}}{\sqrt{2} \, 3^{1/4}
  \pi} \equiv \alpha < 1
\ee
One requires that the expansion of any physical quantity such as energy or
entropy density should have this property. 

Note that including higher order terms does not extend the regime of validity of
the hydrodynamic expansion, but rather improves the accuracy within the
hydrodynamic window. Moreover, in a general boost-invariant dynamical situation 
$\tau_{min}$ does not coincide with thermalization time, since the
non-hydrodynamic (exponential) modes do not have to be negligible at that time
\cite{Chesler:2009cy}.

\subsection{Holographic description of Bjorken flow}

Since the energy-momentum of the
gauge theory is related by AdS/CFT dictionary to the five-dimensional
asymptotically AdS metric, three independent warp factors are to be
expected on the gravity side. This leads to a boost-invariant metric
ansatz of the form
\begin{equation}
\mathrm{d} s^{2} = 2 \hat{g}\left( \ttau, r \right) \mathrm{d} \ttau
\mathrm{d} r + \hat{h} \left( \ttau, r \right) \mathrm{d} r^{2} - r^{2}
\hat{A}\left(\ttau, r\right) \mathrm{d} \ttau^{2} + \left( 1 + r \ttau
\right)^{2} e^{\hat{b} \left( \ttau, r \right)} \mathrm{d} y^{2} +r^{2}
e^{\hat{c}\left( \ttau, r \right)} \mathrm{d} x_{\perp}^{2} \label{bim}
\end{equation}
The functions $g\left( \ttau, r \right)$ and $h \left( \ttau, r \right)$
are gauge degrees of freedom and can be chosen conveniently. Setting
$g\left( \ttau, r \right)$ to 0 and $h \left( \ttau, r \right)$ to
$1/r^{2}$ leads to the Fefferman-Graham-like coordinates, which are
particularly useful in obtaining the boundary energy-momentum tensor. These
coordinates suffer however from an important drawback: they break down at
the locus where $A\left(\ttau, r \right)$ goes to zero, which is 
where one would naively expect the location of the horizon. In order to
keep a full control over the geometry in this region 
it is useful to adopt so called ingoing Eddington - Finkelstein coordinates
which give the following metric ansatz \cite{Heller:2009zz}
\bel{efmetric}
\mathrm{d} s^{2} = 2 \mathrm{d} \ttau \mathrm{d} r - r^{2} A\left(\ttau,
r\right) \mathrm{d} \ttau^{2} + \left( 1 + r \ttau \right)^{2} e^{b \left(
  \ttau, r \right)} \mathrm{d} y^{2} +r^{2} e^{c\left( \ttau, r \right)}
\mathrm{d} x_{\perp}^{2}
\ee
This form of the metric still has a residual gauge freedom $r \rightarrow r + 
R\left( \tau \right)$  \cite{Kinoshita:2008dq}. In the following this freedom
will usually 
not be fixed, since it provides a useful cross-check on the correctness of the
calculations -- physical quantities in the boundary theory should not depend on
the choice of gauge in the bulk. 

The metric \rf{efmetric} describes the gravity dual to
boost-invariant plasma for any $\tau$. It turns out however, that in the
regime of large proper time, the boundary energy density $\epsilon\left(
\tau \right)$ satisfies the equations of hydrodynamics. This highly
nontrivial observation was first made in \cite{Janik:2005zt} and then developed in
\cite{Nakamura:2006ih,Janik:2006ft} using the language of gravity dual. 

It is particularly easy to understand
$\tau^{-2/3}$ damping of the subleading pieces in the expression of the
temperature -- they correspond to the gradient expansion. The gradient
expansion has to have its counterpart in the gravity dual to
boost-invariant flow in the regime of large proper time. The boundary
gradient expansion leads to the introduction of a scaling variable
and the large proper time expansion in the scaling limit (keeping $v = r \cdot 
\ttau^{1/3}$ fixed while taking the limit $\ttau \rightarrow \infty$) on 
the gravity side
\beal{gradexp}
A\left(\ttau, r\right) &=& A_{0} \left( r \, \ttau^{1/3} \right) +
\frac{1}{\ttau^{2/3}} A_{1} \left( r \, \ttau^{1/3} \right) +
\frac{1}{\ttau^{4/3}} A_{2} \left( r \, \ttau^{1/3} \right) + \ldots
\mathrm{,} \nonumber\\ b\left(\ttau, r\right) &=& b_{0} \left( r \,
\ttau^{1/3} \right) + \frac{1}{\ttau^{2/3}} b_{1} \left( r \, \ttau^{1/3}
\right) + \frac{1}{\ttau^{4/3}} b_{2} \left( r \, \ttau^{1/3} \right) +
\ldots \mathrm{,} \nonumber\\ c\left(\ttau, r\right) &=& c_{0} \left( r \,
\ttau^{1/3} \right) + \frac{1}{\ttau^{2/3}} c_{1} \left( r \, \ttau^{1/3}
\right) + \frac{1}{\ttau^{4/3}} c_{2} \left( r \, \ttau^{1/3} \right) +
\ldots \mathrm{.}
\eea
The meaning of the scaling limit is that the radial distance in AdS is measured
in units of inverse temperature ($r / T \sim r \cdot \tau^{-1/3}$ since $T
\sim \tau^{-1/3}$ in the leading order). The large proper time
expansion corresponds to the gradient expansion in the hydrodynamics. The
metric with $A_{0} \left( r \, \ttau^{1/3} \right)$ only will be called the
zeroth order metric and is dual to perfect fluid hydrodynamics on the
boundary side. The first order gravity solution -- $A_{1} \left( r \,
\ttau^{1/3} \right)$, $b_{1} \left( r \, \ttau^{1/3} \right)$ and $c_{1}
\left( r \, \ttau^{1/3} \right)$ -- mimics boundary viscous effects on the
gravity side and is the leading order relevant for the gravitational
entropy production, whereas the second order metric corresponds to
relaxation time in the boundary theory. The formulae providing the form of the  
warp factors at zeroth, first and second orders are given below,
whereas third order formulae can be found in \cite{Math}.

At order zero one has
\begin{eqnarray}
A_{0} \left( v \right) &=& 1-\frac{\pi ^4 \Lambda ^4}{v^4} \mathrm{,} \nonumber \\
b_{0}  \left( v \right) &=& 0 \mathrm{,} \nonumber \\
c_{0} \left( v \right) &=& 0 \mathrm{.}
\end{eqnarray}
which is equivalent to the Janik-Peschanski solution \cite{Janik:2005zt}. For simplicity, 
this solution assumes a choice of gauge to bring out the similarity to the
static AdS-Schwarzschild black brane (see section \ref{revisit}). 

At first order one finds 
\begin{eqnarray}
A_{1} \left( v \right) &=& \frac{2 \pi ^3 (v+\pi \Lambda ) \Lambda ^3}{3
  v^5}+\frac{2 \left(v^4+\pi ^4 \Lambda ^4\right) \delta _1}{3 v^5}
\mathrm{,} \nonumber \\ b_{1} \left( v \right) &=& -\frac{2 \,
  \mathrm{arctan}\left(\frac{v}{\pi \Lambda }\right)}{3 \pi \Lambda
}-\frac{4 \log (v)}{3 \pi \Lambda }+\frac{\log \left(\frac{v^2}{\pi ^2
    \Lambda ^2}+1\right)}{3 \pi \Lambda }+\frac{2 \log \left(\frac{v}{\pi
    \Lambda }+1\right)}{3 \pi \Lambda } \nonumber \\ &&-\frac{4 \log
  \left(\frac{1}{\pi \Lambda }\right)}{3 \pi \Lambda }+\frac{2 \delta _1}{3
  v}-\frac{4}{3 v}+\frac{1}{3 \Lambda } \mathrm{,} \nonumber \\ c_{1}
\left( v \right) &=& - \frac{1}{2} b_{1} \left( v \right) + \frac{\delta
  _1}{v} \mathrm{.}
\end{eqnarray}

The second order formulas can be found in appendix \ref{appb}.

The zeroth order solution looks like the boosted and
dilated black brane. This is in line with the argument of \cite{Bhattacharyya:2008jc}, which
states that the gravity dual to the general solution of hydrodynamics can
be obtained by boosting and dilating black brane and having this as leading
order solution solving Einstein equation perturbatively in the gradient
expansion. Since boost-invariant flow is a particular solution of the
equations of hydrodynamics, this is not a coincidence. The interesting, yet
not surprising given its interpretation as one-dimensional expansion,
feature of boost-invariant hydrodynamics is that the temperature does not
settle down to the constant non-zero value, which gives rise to the  question as to whether this
solution is covered by an event horizon. In the next sections it will be
demonstrated that this indeed the case, which means that the gravity dual
to boost-invariant flow is a well defined geometry from the cosmic
censorship hypothesis point of view.

The gravity solution displayed above determines, via holographic
renormalization, the boundary energy momentum tensor \cite{Balasubramanian:1999re}, which is of the form
\rf{BIstresstensor}. Explicitly, one finds
\begin{align}
\label{edensgrav}
\epsilon\left(\tau\right) &=  \frac{3}{8} N_c^2 \pi ^2 \frac{\Lambda
  ^4}{\tau^{4/3}} 
\left\{1-\frac{2}{3 \pi  \Lambda} \cdot \frac{1}{\tau
^{2/3}}+\frac{1+2 \log (2)}{18 \pi ^2 \Lambda ^2} \cdot \frac{1}{\tau
^{4/3}} + \nonumber \right. \\
&+ \left. \frac{-3+2 \pi ^2+24 \log (2)-24 \log ^2(2)}{486 \pi ^3
\Lambda ^3 \tau ^2} + \ldots\right\}
\end{align}
Furthermore, the proper-time dependence is exactly what is
required by hydrodynamics. The temperature (determined by 
\rf{edensgrav} and \rf{edens}) reads
\begin{align}
\label{tempgrav}
T\left(\tau\right) &= \frac{\Lambda}{\tau^{1/3}} \left\{ 1-\frac{1}{6
\pi  \Lambda} \cdot \frac{1}{\tau^{2/3}}+\frac{-1+\log (2)}{36 \pi ^2
\Lambda ^2} \cdot \frac{1}{\tau^{4/3}}+ \nonumber \right. \\
&+ \left. \frac{-21+2 \pi ^2+51 \log
(2)-24 \log ^2(2)}{1944 \pi ^3 \Lambda ^3} \cdot \frac{1}{\tau^2}
\right\}
\end{align}
Comparison with the static situation determines the coefficient $e_0$
in \rf{edens} to be \cite{Benincasa:2007tp}: 
\be
e_0 = \frac{3}{8} N_{c}^{2} \pi^2 \ .
\ee
Matching the energy density with \rf{edens} and 
\rf{hydrotemp} and using the linearized hydrodynamics to obtain the relaxation time from sound wave dispersion relation
determines the transport coefficients of N=4 Yang-Mills plasma:
\bea
\eta &=& \frac{1}{4 \pi} s\\
\tau_{\Pi} &=& \frac{2 - \log{2}}{2 \pi T} \nonumber \\
\lambda_{1}  &=& \frac{s}{8 \pi^{2} T}
\eea
These expressions are based on terms up to second order. The formulae
\rf{edensgrav} and \rf{tempgrav} include third
order terms, but since the tensorial structure of the third order 
hydrodynamics has not been yet investigated
\footnote{It seems that there is no
practical need to do so -- the inclusion of second order hydrodynamic terms in
RHIC simulations changes the results by just couple of percent
\protect \cite{Luzum:2008cw}. The third order terms will give even more suppressed
contribution. However there are hints that in some cases resummation of
hydrodynamic might be needed \protect \cite{Lublinsky:2009kv}.}
, the relevant transport
coefficients cannot be calculated.

\section{Horizons in the boost-invariant spacetime}
\label{horbi}

\subsection{Preliminaries}

With the tools from the last section in hand one can now turn to the
identification and study of various notions of horizons in the spacetime defined by the
bulk metric \rf{efmetric}. Since the main goal is to study possible notions of
entropy in the dual field theory, the hypersurfaces of interest are those which
satisfy the symmetries of the boundary dynamics under consideration (Bjorken
flow). This singles out spacelike three-surfaces of constant $\tau$ and
$r$. Such surfaces possess outward and inward pointing null normals
\bea
\ell^a &=&  \left[ \frac{\partial}{\partial \tau} \right]^a + 
\half r^2 A(\tau, r) \left[\frac{\partial}{\partial r} \right]^a 
 \ , \nonumber \\  
n^a &=&  - \left[\frac{\partial}{\partial r} \right]^a  \ . \label{nullscaling}
\eea
As usual, these vectors have been cross-normalized so that $\ell \cdot n =
-1$. The rescaling freedom of $\ell \rightarrow f(\tau) \ell$ and $n \rightarrow
n/f(\tau)$ remains but the specific 
scaling used here is chosen to be consistent with the flow of time at asymptotic
infinity. This will be discussed  
in more detail below.  

The hypersurfaces of interest will all lie within the $\tau = \mbox{constant}$
hypersurfaces. Given that coefficients of the metric are given by the series 
\rf{gradexp}, solutions can be sought in the form 
\beal{gensurf}
r_S(\tilde{\tau}) = \frac{1}{\tilde{\tau}^{1/3}} (r_0 +
\frac{1}{\tilde{\tau}^{2/3}} r_1 + 
\frac{1}{\tilde{\tau}^{4/3}} r_2  + r_3 \frac{1}{\tilde{\tau}^{2}}\dots) \, .
\eea
The coefficients $r_k$ appearing here will be determined by the conditions
imposed, and it turns out that the solutions are unique. 
Not only is the event
horizon uniquely defined, but in this case demanding that the FOTH shares
the symmetries of the spacetime means that we can also select a \emph{unique}
FOTH. 
In this section the focus will be on the FOTH and the event horizon,
while in the following section more general surfaces of the form \rf{gensurf}
will play an essential role.

\subsection{The boost-invariant FOTH}

To find
marginally trapped three-surfaces within the $\tau = \mbox{constant}$
hypersurfaces, one needs to 
solve the equation $\tl = 0$. Evaluating $\tl$ on a hypersurface of the form
\rf{gensurf} yields 
\be
\theta_{(\ell)} = \frac{1}{\tau^{1/3}}
\left\{
\frac{3}{2} r_{0} \left(1-\frac{\pi ^4 \Lambda ^4}{r_{0}^4}\right)
+ \frac{1}{\tau^{2/3}} \, \frac{3 \pi ^4 \left(3 r_1+\delta _1+1\right) \Lambda 
  ^4+2 \pi ^3 r_0 \Lambda 
  ^3+r_0^4 \left(3 r_1+\delta _1+1\right)}{2 r_0^4}
\right\} \, . 
\ee
Solving $\theta_{(\ell)} = 0$ (to third order) shows that there is 
a unique apparent horizon on the $\tau=\mbox{constant}$ slices at:
\beal{rah}
r_{AH} (\tau)&=& 
\frac{1}{\tau^{1/3}} \left\{
\pi  \Lambda + 
\frac{1}{\tau^{2/3}} \left(-\frac{\delta_1}{3}-\frac{1}{2}\right) + 
\frac{1}{\Lambda \, \tau^{4/3}} \left(\frac{\Lambda \delta_2}{3}+\frac{1}{9
  \pi}-\frac{1}{24}-\frac{\log (2)}{18 \pi}\right) + \right. \nonumber \\
&+& \frac{1}{\Lambda^2 \, \tau ^2} \left(\frac{\mathcal{C}}{18 \pi ^2}+\frac{\Lambda
  \delta_3}{3}-\frac{25}{432 \pi}+\frac{1}{81 \pi ^2}-\frac{1}{7776}+\frac{7 
  \log^2(2)}{162 \pi ^2} + \right. \nonumber \\ 
&-& \left. \left. \frac{\log (\pi  \Lambda )}{18 \pi}-\frac{2
   \log (\pi  \Lambda )}{27 \pi ^2}-\frac{25 \log(2)}{162 \pi ^2}\right)
\right\} \ . 
\eea

For the three-surface defined by $r=r_{AH}(\tau)$, the inward expansion is
\beal{thetan}
\theta_{(n)} &=& \tau^{1/3}
\left\{
-\frac{3}{\pi \Lambda} 
-\frac{1}{2 \pi ^2 \Lambda ^2 \, \tau ^{2/3}}
-\frac{-1+2 \log (2)}{12 \pi ^3\Lambda ^3 \, \tau ^{4/3}}
+ \frac{1}{\Lambda^4 \, \tau^2}
\left(\frac{7 \log ^2(2)}{54 \pi ^4}-\frac{\log (2)}{24 \pi
^3} + \right. \right. \nonumber \\
 &-& \left. \left. \frac{25 \log (2)}{54 \pi ^4}-\frac{1}{2592 \pi
  ^2}+\frac{\mathcal{C}}{6 \pi ^4}+\frac{35}{216 \pi ^4} 
  \right) \right\} \ , \label{tn}
\eea
where $\mathcal{C}$ is Catalan's constant. 
As expected, \rf{thetan} is independent of the constants $\delta_k$ which
reflect the gauge  
dependence of \rf{efmetric}. It is easy to see numerically that
\be
\theta_{(n)} = -\frac{0.95}{\Lambda }-\frac{0.051}{\Lambda ^2 \tau
^{2/3}}-\frac{0.0010}{\Lambda ^3 \tau
  ^{4/3}}-\frac{0.00039}{\Lambda ^4 \tau ^2}
\ee
This is clearly negative and will stay negative in a neighbourhood of
$r_{AH}$. Further it is clear from the  
expression for the outward expansion that for $r \approx r_{AH}$, $r > r_{AH}
\Rightarrow \tl > 0$ and $r < r_{AH} \Rightarrow  
\tl < 0$. That is, there are fully trapped surfaces ``just-inside'' $r=r_{AH}$
and so this marginally trapped surface bounds a fully  
trapped region and so can be identified as a black brane apparent horizon or
FOTH.  

The remaining geometric quantities discussed in section \ref{blabrame} can now
be calculated.  
First, requiring that the evolution vector $\mathcal{V}^a$ be tangent to the
horizon, implies that in  
the large $\tau$ expansion the expansion parameter $C$ has the form 
\be
C = C_{-1} + \frac{1}{\tau^{2/3}} C_0 +\frac{1}{\tau^{4/3}} C_1
+\frac{1}{\tau^{2}} C_2 +\frac{1}{\tau^{8/3}} C_3 + O
\left(\frac{1}{\tau^{11/3}} \right) \ ,
\ee
for some set of coefficients $C_{-1},\dots C_{3}$. In this case it is  
straightforward to see that $C_{-1}=0$ identically 
in consequence of the structure of the large $\tau$ expansion (even
without using the explicit form of the solution). The coefficients $C_{0}$
(perfect fluid) and $C_{1}$ (viscosity) also
turn out to vanish, so the leading contribution appears at order
$\frac{1}{\tau^{2}}$. All in all, on the $r=r_{AH}(\tau)$ FOTH one
finds  
\bea
C = \left( \frac{1}{9} \right) \frac{1}{\tau^2} 
-\frac{1}{\Lambda \tau^{8/3}} \left(\frac{\log (2)}{9 \pi}-\frac{1}{54} \right) 
+ O \left( \frac{1}{\tau^{11/3}}
\right) \, .  
\eea
This is clearly greater than zero and so the horizon is dynamical: that is
spacelike and expanding. This can then be cross-checked 
in two ways. First one can directly calculate the volume element on the
three-surfaces: 
\bea
\tilde{\epsilon} = \sqrt{\tilde{q}} dy \wedge dx_1 \wedge dx_2 \, 
\eea
where, up to third order
\bea
\sqrt{\tilde{q}} &=& 
\pi ^3 \Lambda ^3 
- \frac{1}{\tau^{2/3}}\frac{1}{2} \pi ^2 \Lambda ^2
+ \frac{\Lambda}{\tau^{4/3}} \left(\frac{1}{4} \pi  \log (2) +\frac{\pi ^2
}{24}+\frac{\pi}{12}\right)   \\ 
&-& \frac{1}{\tau^{2}} \left(\frac{5}{216}-\frac{\pi }{144}-\frac{\pi ^2}{2592}+\frac{5 \log
(2)}{216}-\frac{1}{24} \pi  \log (2)-\frac{35 \log
  ^2(2)}{216}\right) \nonumber
\eea
Then, to lowest order, the rate of expansion is
\be
\frac{1}{\sqrt{\tilde{q}}} \frac{d \sqrt{\tilde{q}}}{d\tau} 
 = \frac{1}{\tau^{5/3}} \left\{
\frac{1}{3 \pi  \Lambda }
+ \frac{1}{\Lambda^2 \, \tau^{2/3}} \left(
-\frac{\log (2)}{3 \pi ^2}-\frac{1}{18 \pi}+\frac{1}{18 \pi ^2} \right)
\right\}
\ee
which is clearly positive. Alternatively
\bea
\frac{1}{\sqrt{\tilde{q}}} \frac{d \sqrt{\tilde{q}}}{d\tau} =
\frac{1}{\sqrt{\tilde{q}}} \Lie_\cV \sqrt{\tq} = - C \tn  \, , \label{Expansion}
\eea
and substituting in the appropriate values obtains the same result.  

Other quantities are the squares of shears in the two null directions:
\bea
\sigma^{(\ell)}_{ab} \sigma_{(\ell)}^{ab} &=& 
\frac{2}{3} \frac{1}{\tau^{2}} 
+ \frac{1}{\Lambda \, \tau^{8/3}}\left(-\frac{2 \log (2)}{3
  \pi}-\frac{1}{9}\right)  \\ 
&+& \frac{1}{\Lambda^2 \, \tau^{10/3}} \left(
\frac{35 \log ^2(2)}{54 \pi ^2}+\frac{\log (2)}{6 \pi}-\frac{14 \log (2)}{27
  \pi^2}-\frac{1}{27 \pi}+\frac{1}{9 \pi ^2}+\frac{1}{648} \right) \nonumber \\ 
\sigma^{(n)}_{ab} \sigma_{(n)}^{ab} & = & 
\frac{1}{\tau^{2/3}} \frac{3}{8 \pi ^4 \Lambda ^4} 
+ \frac{1}{\Lambda^5 \tau^{4/3}} \left(-\frac{3 \log (2)}{8 \pi ^5}-\frac{1}{16
  \pi ^4}+\frac{1}{2 \pi ^5}\right)   \\ 
&+&\frac{1}{\Lambda^6 \tau^{2}} \left(\frac{35 \log ^2(2)}{96 \pi ^6}+\frac{3
  \log (2)}{32 \pi ^5}-\frac{5 \log (2)}{6 \pi ^6}+\frac{1}{1152 \pi
  ^4}-\frac{1}{8 \pi ^5}+\frac{25}{48 \pi ^6} \right) \nonumber
\eea
and the accelerations/inaffinities in the two null directions:
\bea
\kappa_{(\ell)} &=& 
- n_a \ell^b \nabla_b \ell^a = \frac{1}{\tau^{1/3}} \left\{
2 \pi  \Lambda
-\frac{2}{3} \frac{1}{\tau^{2/3}} 
+ \frac{1}{\Lambda \, \tau^{1/3}} \left(\frac{1}{9 \pi}-\frac{\log (2)}{9
  \pi}\right) \right.  \label{kappal} \\ 
&+& \left. \frac{1}{\Lambda^2 \, \tau^{4/3}} \left(
\frac{7 \log ^2(2)}{81 \pi ^2}-\frac{\log (2)}{36 \pi}-\frac{7 \log (2)}{81 \pi
  ^2}+\frac{1}{54 \pi}+\frac{C}{9 \pi ^2}+\frac{1}{162 \pi ^2}-\frac{1}{3888} \right) 
\right\} \nonumber  \\
\kappa_{(n)} &=& - \ell_a n^b \nabla_b n^a = 0 \, \label{kappan0}
\eea
Since $\kappa_{(n)}$ vanishes we have 
\bea
\kappa_{(\cV)} = \kappa_{(\ell)} - C \kappa_{(n)} = \kappa_{(\ell)}\, . \label{kappa0}
\eea
Finally, the connection on the normal bundle is 
\bea
\tom_{a} = - \tilde{q}_a^b n_c \nabla_b \ell^c = 0 \, , \label{tom0}
\eea
and the (three-dimensional) Ricci scalar of the $\tau = \mbox{constant}$ and $v
= v_{AH}$ three-surfaces is 
\bea
\tilde{R} = 0 \, . 
\eea

\subsection{Boost-invariant flow and slow evolution}

Having described the geometry of the FOTH in the boost-invariant
spacetime, the next step is to try to understand its physics. Given that the
metric \rf{bim} is a perturbation of a boosted black brane and that a boosted
brane is simply a coordinate transformation of a static black brane spacetime,
one would intuitively expact that the horizon should be in a near-equilibrium
state. Thus the 
physics of these horizons should be quasi-equilibrium physics. Quantifying this
intuition and then understanding its implications will be the subject of this
section.

One can easily check that these conditions hold for the FOTH in the
boost invariant black brane spacetimes. First, 
\bea
\epsilon^2 \approx {2 \pi ^2 \Lambda ^2} \frac{1}{\tau^{4/3}} \left(1 +
O\left(\frac{1}{\tau^{2/3}}\right)\right) 
\eea
and the evolution parameter for the horizon is of the same order as the
expansion parameter for the metric.  
Next (ii) and (iii) are trivially seen to hold: there is no dependence of
the geometry or scaling of the null normals on $(w,x,y)$ and all 
of these quantities vanish. Thus the horizon is slowly expanding to order
$1/\tau^{2/3}$. Though it is a consequence of these conditions, in this case we can also
explicitly check that the horizon is almost null in the sense of Eq.~(\ref{Ltq_sig}); 
the evolution of the three-slices is characterized by the expansion
and shear associated with $\ell^a$, just as they would be for a  
truly null surface.

One can now check these conditions (and their consequences) for the null scaling 
(\ref{nullscaling}). Though it is clear 
that the conditions will hold for a variety of scalings, we make this particular choice so that
``time'' evolution on the horizon is consistent
with that on the boundary (both are proportional to $\tau$) and, as a retroactive justification, the surface gravity
matches the temperature on CFT side.  
Explicitly checking the conditions, the first part of  (i) holds trivially since
$\tilde{\omega}_a = 0$ by equation (\ref{tom0}) while 
\bea
\kappa_{\cV} = - \cV^a n_b \nabla_a \ell^b = \kappa_{(\ell)}
\eea
by equation (\ref{kappan0}) and so from (\ref{kappal}) the second part holds as well:
\bea
\Lie_{\cV} \kappa_{\cV} \approx - \frac{2}{3} \pi\Lambda \frac{1}{\tau^{4/3}}
\,\Longrightarrow |\Lie_{\cV} \kappa_{\cV}| \lesssim \epsilon^2 \,  .  
\eea
Finally for (ii) a similarly trival calculation from (\ref{tn}) shows that
\bea
\Lie_{\cV} \tn \approx - \frac{1}{\pi\Lambda} \frac{1}{\tau^{2/3}} 
\,\Longrightarrow 
|\Lie_{\cV} \tn| \lesssim \epsilon \, .  
\eea
Thus, the horizon is slowly evolving to order $\epsilon \sim
\tau^{-2/3}$. Then, in addition to it being geometrically ``almost'' null it is
also mechanically close to equilibrium. As already noted, Einstein's equations
are sufficient to imply that the zeroth law of black hole mechanics (almost)
holds: the surface gravity $\kappa_{\cV}$ is approximately constant on each
slice. Here, the symmetry of the black brane means that a stronger result is
valid -- from (\ref{kappal}) one can see that the surface gravity is constant on
each 
slice (though it evolves slowly up the horizon). The dynamical first law also
automatically holds, but again in this case one can also check it explicitly. To 
order $\tau^{-2}$ one has:  
\beal{deltas}
\kappa_{\cV} \Lie_{\cV} \sqrt{\tilde{q}} \approx \sqrt{\tilde{q}} ||
\sigma^{(\ell)} ||^2 \approx \frac{2}{3} \pi^3 \Lambda^3 \frac{1}{\tau^2} \,
, \label{first} 
\eea
and so, as expected, the expansion is driven by the null shears (which would 
normally suggest incoming gravitational radiation). 
The dependence on $\tau$ in this expression
matches what is expected in leading order \footnote{Obviously leading order 
  means first order in the gradient expansion, since entropy is preserved
  in the perfect fluid case.} on the basis of thermodynamics of Yang-Mills 
plasma. To see this, note that the volume in the boost invariant setting
depends linearly on $\tau$, so that the entropy of the plasma can be
written as $S = s \tau V_0$ where $V_0$ is a reference volume and $s$ is
the entropy density \rf{entrodens}.  Using the results of the previous section one then finds 
\be
T \frac{dS}{d\tau} \sim \frac{1}{\tau^2}
\ee
which is consistent with \rf{deltas}.

\subsection{Event horizon}
\label{EH}

It is usually said that the event horizon is somewhat inconvenient to work
with, since determining it requires knowing the entire future evolution
of the spacetime under consideration. This is indeed an onerous requirement
in the typical situation of computing the evolution of spacetime geometry
starting from some initial data. The setting explored in this paper is in a
sense complementary: the spacetime geometry is constructed order by order
in a large proper-time expansion (or gradient expansion) starting in the
far future at zeroth order. This circumstance makes it possible to
determine the location of the event horizon in the late time regime.

The method of finding the event horizon for boost-invariant flow closely
resembles the one presented in \cite{Bhattacharyya:2008xc} for the gravity
duals to fluid dynamics \cite{Bhattacharyya:2008jc}. The crucial assumption
there was that the metric relaxes to (uniformly boosted) AdS Schwarzschild, where the position
of the event horizon is well known. The event horizon for the metrics there
was defined as a unique null surface which asymptotically coincides with
the event horizon of the static AdS Schwarzschild dual to the uniform flow at 
constant temperature. Despite the fact that this is not the case for 
boost-invariant 
flow, it is still possible to find a unique null surface which can be
interpreted as an event horizon. The key observation is that the event
horizon should coincide with the FOTH in the large-proper time regime and
within the scaling limit\footnote{The scaling limit insures that the higher
  derivative corrections to the geometry are small (MH thanks Toby Wiseman
  for discussion on this point)}.  Its radial position in AdS should depend
on proper time only, which reflects the boost-invariance (no rapidity
dependence) together with translational and rotational symmetry in the
perpendicular directions (no $\vec{x}_{\perp}$ dependence). If $r$ is the 
radial direction in AdS space, $\tau$ the proper time and $r_{EH}\left(
\tau \right)$ expresses the time evolution of the horizon, then the
equation defining the sought co-dimention one surface in AdS takes the form
\be
r - r_{EH} \left( \tau \right) = 0\mathrm{.}
\ee
The covector normal to the surface is $dr - r_{EH}' \left( \tau \right)
d\tau$ and requiring it is null gives the equation for $r_{EH} \left(\tau\right)$
\begin{equation}
A\left( \tau, r_{EH} \right) \cdot r_{EH}^{2} - 2 r_{EH} ' = 0 \mathrm{,}
\end{equation}
where for clarity the dependence of $r_{EH}$ on $\tau$ is ommited. This
equation can be solved perturbatively in the scaling limit. 

Using the solution described in \cite{Heller:2009zz} (valid up to third order in
the late proper-time expansion) one finds
\begin{align}
\label{reh}
r_{EH} &= \frac{1}{\tau^{1/3}} \left\{ \pi \Lambda - \left(\frac{1}{2} 
+ \frac{\delta_{1}}{3}\right) \cdot \frac{1}{\tau^{2/3}}
+ \left(  \frac{\delta_{2}}{3}+\frac{1}{6 \pi  \Lambda }-\frac{1}{24
\Lambda }-\frac{\log (2)}{18 \pi  \Lambda } \right) \cdot
\frac{1}{\tau^{4/3}} + \nonumber \right. \\
&+ \left( \frac{\delta_{3}}{3}-\frac{\log
(\Lambda )}{18 \pi  \Lambda ^2}-\frac{2 \log (\Lambda )}{27 \pi ^2
\Lambda^2}-\frac{29}{432 \pi  \Lambda ^2}+\frac{C}{18 \pi ^2 \Lambda
^2}-\frac{5}{324 \pi ^2 \Lambda ^2}-\frac{1}{7776 \Lambda ^2} -
\right. \nonumber \\ 
&- \left. \left. \frac{\log (\pi )}{18 \pi  \Lambda ^2}-\frac{2 \log(\pi )}{27 \pi ^2
  \Lambda ^2}+\frac{7 \log ^2(2)}{162 \pi ^2 \Lambda ^2}-\frac{17 \log
  (2)}{81 \pi ^2 \Lambda 
^2} \right)  \cdot \frac{1}{\tau^{2}} \right\} \mathrm{.}
\end{align}
The dependence on arbitrary constants of integration reflects the gauge 
freedom -- the position of the horizon is a gauge-dependent quantity ($r
\rightarrow r + f\left( \tau \right)$ for arbitrary $f\left( \tau
\right)$). Comparing \rf{rah} with \rf{reh} it turns out that the FOTH coincides
with the event horizon in 
the leading and first subleading orders. This is in agreement with the
observation that the constant $C$ is non-zero only in the second and
higher orders in $\frac{1}{\tau^{2/3}}$ expansion. The second orders differ
and the FOTH becomes spatial. 

It is be also interesting to compare the position of the apparent
and event horizon with the naive horizon defined as the hypersurface
on which $A(\tau,r)$ vanishes
\begin{equation}
r_{\mathrm{naive}} = \frac{1}{\tau^{1/3}} \left\{
\pi \Lambda - \frac{1}{\tau^{2/3}} \left( \frac{1}{3} +
\frac{\delta_{1}}{3} \right) + \ldots
\right\} \ .
\end{equation}
Such a hypersurface coincides with the event and apparent horizon in the
leading order, but differs in the first subleading order -- it is situated
between the event horizon and the boundary. This means, for example, that truncating
numerical simulations at the point where $A(\tau,r)$ vanishes is simply
wrong.

Finally, observe that in the naive limit $\tau \rightarrow \infty$ the
boost-invariant metric relaxes to the empty AdS$_{5}$ metric instead of the
static AdS-Schwarzschild solution. However, this is not so strange from the dual
CFT point of view, where the fluid is expanding to infinity and its energy
density becomes smaller and smaller. It means that the boundary system does
not permanently thermalize to non-zero temperature. The interesting feature of
the boost-invariant flow is an apparent thermalization, which expresses itself
as an applicability of the equations of hydrodynamics in the late stages of the
evolution.

\subsection{Revisiting the scaling limit}
\label{revisit}

The symmetries of boost-invariant flow make it possible to seek the
location of the event horizon considering only the variables $r$ and
$\tau$. It is possible then to focus only on the $dr, d\tau$ part of the
full metric, which at leading order takes the form
\be
ds^2 = 2 d\tau dr - r^2
\left\{1-\frac{\pi^4\Lambda^4}{(r\tau^{1/3})^4}\right\} d\tau^2 
+ \dots 
\ee
The scaling limit discussed at length in section \ref{gradubi} involved
introducing the scaling variable $v=r\tau^{1/3}$ which is kept fixed as
$\tau\rightarrow\infty$. This motivates the following change of variables 
\bea
\tau &=& \left(\frac{2u}{3}\right)^{3/2}\ , \nonumber \\
r &=& \sqrt{\frac{3}{2u}} v\ ,
\eea
which leads to
\be
ds^2 = 2 du dv - v^2 \left\{1-\frac{\pi^4\Lambda^4}{v^4}\right\} du^2 
+ O\left(\frac{1}{u} \right) \dots
\ee
This shows that this part of the metric looks precisely the same as the
corresponding part of the static black brane metric, with $v$ denoting the
radial coordinate and $u$ Eddington-Finkelstein ingoing time
coordinate. This means that at leading order in the late-time expansion the
problem of determining radial 
geodesics\footnote{As stressed previously, this is all that is needed to
  determine the location of the event horizon.} in the 
asymptotic boost-invariant geometry is the same as in the static case. 
It is then not surprising that the naive position of the horizon coincides
asymptotically with the actual event horizon. 

Note however that these considerations do not imply that the asymptotic
geometry is static. Clearly, the remaining terms in the metric are
time-dependent even after this coordinate transformation, even though the
area of the event horizon remains constant in this regime.

\section{Phenomenological notions of entropy}

\subsection{Introduction}

It is believed that equilibrium states of black holes are thermodynamic in
nature. Their entropy is associated with the area of spacelike slices
of the event horizon in an unambiguous way and the second law of thermodynamics is
linked with Raychaudhuri's equation (\ref{RaychaudhuriEqns}). The property that the area of the
event horizon is non-decreasing continues to hold in a generic dynamical
setting. This prompts the question whether there is a sensible notion of entropy
valid in such a non-equilibrium situation. However, as discussed in section
(\ref{AlternateHOR}), there are hypersurfaces of non-decreasing area other than the event
horizon (which coincide with it in the static case). The notion of entropy thus
becomes less clear in these cases, as frequently discussed in the literature. The AdS/CFT correspondence makes it possible to view this problem from the gauge
theory perspective. For example, the teleological nature of the event horizon
leads to acausal behaviour of gauge theory entropy associated with it
\cite{Chesler:2008hg}.  

The bulk description is under control precisely when the field theory is
strongly coupled, which in itself makes it hard to analyse directly. However it
is well understood how to formulate a hydrodynamic description of the system in
the appropriate regime. This description depends only on symmetries and the idea
of the gradient expansion, as explained in \cite{Baier:2007ix,Bhattacharyya:2008jc} for the conformal case and
in \cite{Romatschke:2009kr} in general.

\subsection{Entropy from second order hydrodynamics\label{hydroentrCONS}}
 
The requirement that entropy should be non-decreasing during hydrodynamic
evolution can be expressed in a covariant way in terms of an entropy current
whose divergence is non-negative \cite{LL}.  While the energy momentum tensor is
a canonically defined operator, the entropy current is a derived notion.  In the
spirit of hydrodynamics (or effective field theory) it is also constructed in a
gradient expansion as the sum of all possible terms at a given order.  The
dynamical equations of hydrodynamics are the conservation equations for the
expectation value of the energy-momentum tensor. Thus, the coefficients
appearing in the gradient expansion of the expectation value of the energy-momentum
tensor (the transport coefficients) are the physical parameters of this
phenomenological theory, since they figure directly in the evolution
equations. They describe physical properties of the underlying
quantum field theory. In contrast, the coefficients which appear in the
phenomenological expression for the entropy current are constrained only
by the requirement that its divergence be non-negative. These parameters are
logically independent of the transport 
coefficients. At the present level of understanding they reflect a real
ambiguity in the phenomenological notion of entropy current in hydrodynamics (as
explained in the following subsection). This ambiguity is however of no
consequence when entropy differences between equilibrium states are considered.

In the case of conformal fluids the most
general form of the entropy current was recently constructed
\cite{Bhattacharyya:2008xc,Romatschke:2009kr} up to 
second order in gradients. The crucial symmetry requirement was Weyl
covariance (see \cite{Loganayagam:2008is} for useful explicitly Weyl-covariant formulation of hydrodynamics). Using the notation of \cite{Romatschke:2009kr} (which descends from
\cite{Bhattacharyya:2008xc}) one has\footnote{In the hydrodynamic formulas
  $\Delta^{\mu \nu} = \eta^{\mu \nu} + u^{\mu} u^{\nu}$ is projector to the
  fluid's local rest frame, whereas $\nabla_{\perp}^{\mu} = \Delta^{\mu \nu}
  \nabla_{\nu}$. Moreover fluid shear tensor (responsible for dissipation in the
  first order conformal hydrodynamics) reads $\sigma^{\mu \nu} = \Delta^{\mu
    \alpha} \Delta^{\nu \beta} \left( \nabla_{\alpha} u_{\beta} + \nabla_{\beta}
  u_{\alpha} \right) - \frac{1}{4-1} \Delta^{\mu \nu} \nabla_{\alpha}u^{\alpha}$
  and vorticity (nonzero for rotating fluid) $\Omega^{\mu \nu} = \Delta^{\mu
    \alpha} \Delta^{\nu \beta} \left( \nabla_{\alpha} u_{\beta} - \nabla_{\beta}
  u_{\alpha} \right)$.} 
\beal{gencur}
S^{\mu}_{\rm non-eq}&=&s u^\mu+\frac{A_1}{4} {\cal S}_1 u^\mu + A_2 {\cal S}_2 u^\mu + 
A_3\left(4 {\cal S}_3 - \frac{1}{2}{\cal S}_1 +2 {\cal S}_2\right) u^\mu\nonumber\\
&&\hspace*{2.5cm}+ B_1 \left(\frac{1}{2}{\cal V}_1^\mu +\frac{u^\mu}{4} {\cal S}_1\right)+ B_2 
\left({\cal V}_2^\mu-u^\mu {\cal S}_2\right)\,.
\label{ecc}
\eea
Here $s$ denotes the thermodynamic 
entropy density \rf{entrodens}, and ${\cal S}_{1,2,3}$ are the 
3 possible conformal (Weyl-covariant) scalars  
\bea
&{\cal S}_1=\sigma_{\mu \nu}\sigma^{\mu \nu}\,,\quad
{\cal S}_2=\Omega_{\mu \nu}\Omega^{\mu \nu}\,,&\nonumber\\
&{\cal S}_3=c_s^2\npu_\mu \npd^\mu \ln \sn+\frac{c_s^4}{2}\npu_\mu \ln \sn \npd^\mu \ln \sn
-\frac{1}{2}u_\alpha u_\beta R^{\alpha \beta}-\frac{1}{4}R+\frac{1}{6} \mt^2\,,&
\label{confscals}
\eea
and ${\cal V}_{1,2}$ are 2 possible conformal vectors\footnote{Note the absence
  of parity breaking terms present in \cite{Bhattacharyya:2008xc}. For
  discussion of hydrodynamics with parity breaking terms see
  \cite{Son:2009tf}.}
\bea
{\cal V}_1^\mu = \npu_\alpha \sigma^{\alpha \mu}+2 c_s^2 \sigma^{\alpha \mu}\npu_\alpha \ln \sn
-\frac{u^\mu}{2} \sigma_{\alpha \beta}\sigma^{\alpha \beta}\,,\quad
{\cal V}_2^\mu = \npu_\alpha \Omega^{\mu \alpha}+u^\mu \Omega_{\alpha \beta} \Omega^{\alpha \beta}\,,
\label{confvec}
\eea
where $\sigma$ is the hydrodynamic shear tensor and $\Omega$ is the vorticity,
and $R$ is the boundary Ricci tensor. 

The entropy current depends (\ref{gencur}) on 5 constants $A_{1,2,3}$ and
$B_{1,2}$ and its divergence reads 
\bea
\label{entropyc}
\nabla_\mu S^\mu_{\rm non-eq}&=&\frac{1}{2}\npu_{\mu}\npu_\nu \sigma^{\mu \nu}
\left(-2 A_3+ B_1\right)
+\frac{1}{3}\npu_\mu \sigma^{\mu\nu}\npu_\nu \ln \sn \left(-2A_3+ B_1\right)\nonumber\\
&&\hspace*{-2cm}+\sigma_{\mu \nu} \left[\frac{\eta}{2T} \sigma^{\mu \nu}+R^{\mu \nu} \left(
-\frac{\kappa}{2 T}+A_3\right)
+u_\alpha u_\beta R^{\alpha <\mu \nu> \beta}\left(\frac{\kappa-\eta\tau_\pi}{T}
+A_1+B_1-2A_3\right)\right.\nonumber\\
&&\hspace*{-2cm}\left.-\frac{1}{4}\sigma^{\mu}_{\ \lambda} \sigma^{\nu \lambda}\left(
\frac{2\lambda_1-\eta \tau_\pi}{T}
+A_1+B_1-2A_3\right)
+\frac{1}{3}\npd^{<\mu} \npd^{\nu>}\ln \sn\left(\frac{\eta \tau_\pi}{T}-A_1-2A_3\right)
\right.\nonumber\\
&&\hspace*{-2cm}\left.+\Omega^\mu_{\ \alpha} \Omega^{\nu \alpha}
\left(-\frac{\lambda_3+2 \eta \tau_\pi}{2T}+A_1-2A_2-2A_3+B_1\right)
\right.\\ 
&&\hspace*{-2cm}\left.
+\sigma^{\mu \nu}\frac{\mt}{12}\left(\frac{2 \eta \tau_\pi}{T}-2A_1+6 A_3-5 B_1\right)
+\frac{1}{9}\npd^{<\mu} \ln \sn \npd^{\nu>}\ln \sn\left(-\frac{\eta \tau_\pi}{T}+A_1+B_1\right)
\right]\,.\nonumber
\eea
If the shear tensor is non-vanishing\footnote{Note that $\sigma_{\mu \nu}
  \sigma^{\mu \nu}$ as a
  trace of the square of matrix cannot be negative and vanishes if and only if
  $\sigma^{\mu \nu}=0$.} the positivity of the shear viscosity $\eta$
should guarantee (see however \cite{Romatschke:2009kr} and the next footnote) that divergence of the entropy current is non-negative: higher
order terms cannot change this conclusion as long as the gradient expansion is
valid. However, as noted in \cite{Bhattacharyya:2008xc}, it is perfectly
reasonable for $\sigma^{\mu \nu}$ to locally vanish (requiring this imposes just
5 conditions 
for derivatives of the four-velocity) in which case the higher order terms will 
dominate the
entropy production. Positivity of \rf{entropyc} thus requires 
\be
\label{B1eq2A3}
B_1=2 \, A_3 \mathrm{.}
\ee

Since the shear tensor
$\sigma^{\mu \nu}$ is multiplying the whole square bracket in
\rf{entropyc}, in the case when it is 0 
the whole contribution from first two orders is absent. 
At this level there is a
real 4-parameter ambiguity in the hydrodynamic construction of the entropy
current\footnote{Considerations of the case with an arbitrary small 
  $\sigma^{\mu \nu}$ in \cite{Romatschke:2009kr} suggest that further
  constraints on the entropy current may be imposed. In particular, the only
  freedom left after such consideration is in the parameter $A_{1}$. 
  It appears that these arguments rest on competition between terms of different
  orders in the gradient expansion. This paper cannot shed any light on this,
  since in the boost invariant case the shear is not close to vanishing.}.

\subsection{The case of Bjorken flow}

From the perspective of the AdS/CFT correspondence it is natural to ask whether
the ambiguities appearing in the construction of the hydrodynamic entropy
current match on both sides of the duality. In a general situation this might be
involved, but one may try to gain some insight into this question by considering
a particular solution. The Bjorken flow provides a simple, highly symmetric, yet
non-trivial example. 

The current \rf{gencur} evaluated on the boost-invariant solution  
given by the velocity $u^\mu = [\partial_\tau]^\mu$ 
and temperature $T(\tau)$ \rf{hydrotemp} 
takes the form
\bel{curphen}
J^\mu = \tilde{s} u^\mu
\ee
with 
\bel{entrononeq}
\tilde{s}(\tau) = s(T(\tau)) \left\{1 + 2\, \frac{A_1 - A_3 + B_1}{3 \pi ^2
  T(\tau)^2  
  \tau^2}\right\} 
\ee
In general the entropy current does not have
to be proportional to the flow velocity beyond leading order (perfect fluid),
but in the special case of boost-invariant flow non-leading order effects are
captured by the single scalar function $\tilde{s}(\tau)$. 
This function involves 3 of the 5 constants appearing in the 
general phenomenological construction. 
Changing the value of $A_1$ has been identified with the freedom in choosing the
horizon 
to boundary map \cite{Bhattacharyya:2008xc}. The ambiguity parametrized by $A_3$
was not interpreted in \cite{Bhattacharyya:2008xc}. 
One would like to 
interpret this freedom in terms of allowed definitions of ``horizon'' on 
gravity side. Note that the example of Bjorken flow, while rather special, is
still rich enough to partially capture this ambiguity.

In quantitative terms 
this ambiguity can be estimated as follows. In order for the hydrodynamic
expansion 
of \rf{entrononeq} to be
valid the magnitude of $|A_1+A_3|$ should bounded so that the leading term
dominates for times larger than $\tau_{min}$ defined by \rf{taumin}. Expanding
\rf{entrononeq} up 
to second order one has
\be
\tilde{s}(\tau) \sim 
\frac{\Lambda^3}{\tau} \left\{1 
-\frac{1}{2 \pi  \Lambda } \frac{1}{\tau^{2/3}}
+  \frac{ (8 (A_1+A_3) + \log (2))}{12 \pi ^2 \Lambda^2} \frac{1}{\tau^{4/3}}
\right\}
\ee
Demanding that the second order contribution be smaller than the first order
correction by a factor of $\alpha \beta$ at $\tau=\tau_{min}$ leads
to the bound  
\be
\frac{1}{8} (-\beta -\log (2)) < A_1 + A_3 < \frac{1}{8} (\beta -\log (2))
\ee 
where $\beta$ is at most of order $1/\alpha$. This 
provides a rough estimate of the allowed inderminacy in the phenomenological
notion of non-equilibrium entropy as defined by \rf{entrononeq}:
\be
\tilde{s}(\tau) \sim 
\frac{\Lambda^3}{\tau} \left\{1 
-\frac{1}{2 \pi  \Lambda } \frac{1}{\tau^{2/3}}
\pm  \frac{\beta}{12 \pi ^2 \Lambda^2} \frac{1}{\tau^{4/3}}
\right\}
\ee
One would like to understand this quantitively in terms of the freedom of
defining entropy on the gravity side.

\subsection{Entropy from gravity}

As reviewed in section \ref{BBmech}, in a dynamical setting it is no longer clear
if there is an appropriate geometrical notion which can be used for the
definition of 
entropy. 
For example both apparent and event horizons appear to give rise to notions of
entropy, which 
satisfy the second law of thermodynamics and coincide in equilibrium. This
provides motivatation to look more generally at the 
dynamics of hypersurfaces whose area is non-decreasing. 
The starting point should be equation (\ref{AreaExp}). This formula 
determines the rate of change of the area element of a general 3-hypersurface in
the terms of the expansions $\thetan$, $\thetal$ and the expansion parameter $C$.  
In the boost-invariant case the 3-hypersurfaces consistent with the boundary
symmetry have the form \rf{gensurf}. 
Their area is given by
\be
\mathcal{A} = \pi ^3 \Lambda ^3 \left\{ 1 + 
\frac{3 \tilde{r}_1  + 1}{\pi\,\Lambda\,\tilde{\tau}^{2/3} } +
\frac{36 \tilde{r}_1^2 +24 \tilde{r}_1 + 36 \pi  \tilde{r}_2 +2 \pi +5 \log (2)}{12 \pi
  ^2\,\Lambda ^2\,\tilde{\tau}^{4/3}}
\right\}
\ee
where 
\bea
\tilde{r}_1 = r_1 - \third \delta_1 \nonumber \\
\tilde{r}_2 = r_2 + \third \delta_2 
\eea
The notion of entropy defined by such hypersurfaces is 
\bel{entgrav}
S = \frac{N_c^2}{2\pi} \mathcal{A}
\ee
(since in the units used here (AdS radius set to 1) $G_N^{-1}=2 \pi^{-1} N_{c}^{2}$). 

Equation \rf{AreaExp}, which expresses the change of area  of the hypersurface
sections becomes  
\beal{areaexpeval}
\Lie_{\cV} \sqrt{\tq} &=& \sqrt{\tq} (\tl - C \tn)  = \nonumber \\
&=& -\frac{6 \tilde{r}_1 + 2}{3
  \pi  \Lambda } \frac{1}{\tilde{\tau}^{5/3}} +   
\frac{18 \tilde{r}_1^2+12 \tilde{r}_1 - 2 \pi  (18 \tilde{r}_2 + 1)+ 6 - 
5 \log (2)}{9 \pi ^2 \Lambda ^2}  \frac{1}{\tilde{\tau}^{7/3}}
\eea 
In this formula $r_0$ has been set to $\pi$ to match the thermodynamic entropy
when all gradient corrections are discarded. For
the leading term (at order $1/\tilde{\tau}^{5/3}$)  
to be non-negative one gets the bound 
$\tilde{r}_1 < -1/3$, and then requiring that the following term be smaller
gives an allowed 
range for $r_2$. At this level of analysis this is all one gets;  $r_1$ is 
not fixed\footnote{Note in particular that there are surfaces outside the event
  horizon which are acceptable from this point of view.}. Both notions of
entropy (defined via the event or apparent horizon) provide unique 
$r_1$, which lies in the allowed range. 

The entropy density\footnote{Entropy density is understood as entropy per unit
  volume, which in the proper-time -- rapidity coordinates involves a factor of
  $\tau$.} obtained from the third order expression for the event 
horizon reads
\begin{align}
s_{EH} \left( \tilde{\tau} \right) &= \frac{1}{2} N_c^2 \pi ^2 \Lambda ^3
\left\{1-\frac{1}{2 \pi  \Lambda} \cdot
\frac{1}{\tilde{\tau}^{2/3}}+  \frac{6+\pi +6 \log (2)}{24 \pi ^2 \Lambda
^2}\cdot\frac{1}{\tilde{\tau}^{4/3}} - \right. \nonumber \\ 
&- \left. \frac{420+90 \pi +\pi ^2+372 \log (2)+108
\pi  \log (2)+420 \log^2(2)}{2592 \pi^3 \Lambda ^3 \tilde{\tau} ^2}\right\}
\end{align}

In the dynamical case the result for the entropy density
reads
\bea
s_{AH} &=& \frac{1}{2} N_c^2 \pi ^2 \Lambda ^3 
\frac{1}{\tau} \left\{1 - \frac{1}{\tilde{\tau}^{2/3}} \frac{1}{2 \pi  \Lambda }
+ \frac{1}{\Lambda^2 \tilde{\tau}^{4/3}}
\left(\frac{\log (2)}{4 \pi ^2}+\frac{1}{24 \pi}+\frac{1}{12 \pi ^2}\right)
\right. \nonumber \\ 
&-& \left. \frac{1}{\Lambda^3 \tilde{\tau}^{2}} \left(\frac{35 \log ^2(2)}{216 \pi
  ^3}-\frac{\log (2)}{24 \pi ^2}+\frac{5 \log (2)}{216 \pi ^3}-\frac{1}{2592
  \pi}-\frac{1}{144 \pi ^2}-\frac{5}{216 \pi ^3}\right)
\right\}
\eea

Numerically one finds
\beal{entroah}
s_{AH} = \frac{1}{2} N_c^2 \pi ^2 \Lambda ^3 
\frac{1}{\tau} \left\{1
- \frac{1}{\tilde{\tau}^{2/3}} \frac{0.16}{\Lambda } + 
\frac{1}{\tilde{\tau}^{4/3}} \frac{0.039}{\Lambda ^2}
-\frac{1}{\tilde{\tau}^{2}} \frac{0.0065}{\Lambda ^3}
\right\}
\eea
For the event horizon
\beal{entroeh}
s_{EH} = \frac{1}{2} N_c^2 \pi ^2 \Lambda ^3 
\frac{1}{\tau} \left\{1-\frac{0.16}{\Lambda  \tilde{\tau} ^{2/3}}+\frac{0.056}{\Lambda ^2
  \tilde{\tau} 
^{4/3}}-\frac{0.018}{\Lambda ^3 \tilde{\tau} ^2}
\right\}
\eea

The key observation is that the event
and apparent horizons coincide at the leading and first subleading orders, which
is a hint that also in the case of a general surface there should be no
ambiguity until the second subleading 
order. If one is to identify the entropy defined here with the field theory
observable, as required by the AdS/CFT correspondence, then it should be
Weyl covariant in the boundary sense. 
To do this explicitly one would need to solve the Einstein equations with the 
boundary metric given by 
\be
ds_4^2 = e^{-2\omega(\tau)} \left\{-d\tau^2 + \tau^2 dy^2 + dx_\perp^2\right\}
\ee
where $\omega(\tau)$ is a conformal factor having the form of an expansion 
in powers of $1/\tau^{2/3}$. The entropy computed this way would be Weyl
covariant (i.e. proportional to the appropriate power of the conformal factor)
only for $\tilde{r}_1=-1/2$., i.e. the value assumed by $\tilde{r}_1$ in the
case of the event or 
apparent horizon (which coincide at this order). The quick way to get this
answer is to write the entropy (\ref{entgrav}) in terms of temperature and
velocity, whose transformation rules under Weyl rescalings are known. This
procedure parallels the field theory analysis reviewed earlier. The first step
is factor out the thermodynamic entropy which sets the Weyl transformation
property of the entropy density. This leads to  
\beal{entrohor}
s = \half N_c^2 \pi^2 T(\tau)^3 \left\{1 + \frac{6 \tilde{r}_1+3}{2 \pi
  \tau^{2/3} \Lambda } + 
\frac{36 \tilde{r}_1^2 + 42 \tilde{r}_1 + 36 \pi  \tilde{r}_2 + 2 \pi +9+4 \log
  (2)}{12 \pi 
   ^2 \tau^{4/3} \Lambda ^2}\right\}
\eea
Since there are no Weyl-covariant scalars nor vectors at first order in
derivatives, the only way that this formula can be Weyl covariant is if the
first subleading term vanishes, which determines $\tilde{r}_1=-1/2$. When this
result is used in eq. \rf{areaexpeval}, one finds
\beal{areaexphor}
\Lie_{\cV} \sqrt{\tq} &=& \sqrt{\tq} (\tl - C \tn)  = \nonumber \\
&=& \frac{1}{3\pi  \Lambda } \frac{1}{\tilde{\tau}^{5/3}} +   
\frac{9 - 4 \pi  (18 \tilde{r}_2 + 1) - 
10 \log (2)}{18 \pi ^2 \Lambda ^2}  \frac{1}{\tilde{\tau}^{7/3}}
\eea 
The leading contribution ensures positive entropy production due to the shear 
viscosity so there are no further constraints on $\tilde{r}_2$. 
The 
appearance of possible freedom in choosing $\tilde{r}_2$ can be understood
following again the hydrodynamic argument 
presented in subsection (\ref{hydroentrCONS}). In the first place, note that the formula
\rf{areaexphor} is evaluated using the bulk Eddington-Finkelstein proper-time
$\tilde{\tau}$, 
which raises the question about its relation with boundary proper-time
coordinate $\tau$. Such a mapping freedom has been addressed using 
Weyl-covariant language in \cite{Bhattacharyya:2008xc} and amounts to the
trivial mapping in the 
first two orders of the gradient expansion, with an ambiguity showing up at the
second order. In the case of boost-invariant flow the most general mapping (up
to second order) takes the form
\be
\tilde{\tau} \longrightarrow \tau (1 + \frac{\delta A_1}{\Lambda^2\,\tau^{4/3}})
\ee
where $\delta A_1$ is a constant parameter multiplying Weyl-covariant scalar
$S_{1}$. For such mappings to make sense 
within the context of the gradient expansion this parameter must be suitably bounded as
explained earlier. This shifts the second order contribution in \rf{entroah} and
\rf{entroeh}, so it must be of order $\alpha^2$ in the sense of
\rf{taumin}, which leads to
\be
\delta A_1 = \pm \frac{\gamma}{32\pi^2}
\ee
with $\gamma$ of order 1. This mapping freedom accounts only for a part of the
ambiguity in the phenomenological construction of the hydrodynamic entropy
current. The rest of the ambiguity can be understood following the Weyl analysis
of the properties of gravitational entropy. At second order in gradients there
are two Weyl covariant scalars ($S_{1}$ and $S_{3}$) and one vector ($V_{1}$)
which are non-trivial when evaluated on the boost-invariant solution, with the
vector being proportional to velocity in that particular case. Since mapping
freedom is identified with the $S_{1}$ contribution\footnote{It is also
  reasonable to stipulate that $A_1$ is partly linked with the mapping freedom
  and partly with a freedom in choosing between different notions of horizons in the
  bulk.}, it is clear the $\tilde{r}_{2}$ must come from the relevant
combination of $S_{3}$ and $u_{\mu} V^{\mu}$. Evaluating \rf{entrohor} with
$\tilde{r}_1=-1/2$ gives
\bel{entrohor2}
s = \half N_c^2 \pi^2 T(\tau)^3 \left\{1 + \frac{-3 + 36 \pi  \tilde{r}_2 + 2
  \pi + 4 \log (2)}{12 \pi^2 \tau^{4/3} \Lambda^2}\right\} 
\ee
Comparing this with \rf{entrononeq} one finds 
\be
\tilde{r}_2 = \frac{2}{9 \pi }\left(A_1- A_3 + B_1\right) +\frac{1}{12 \pi
   }-\frac{1}{18}-\frac{\log (2)}{9 \pi }
\ee
Although this is all one can get from the analysis of the gravity dual to the
boost-invariant flow\footnote{Note that in hydrodynamics due to the requirement
  of positive divergence of the entropy current, $A_{1}$ and $B_{1}$
  contributions to the entropy current are not independent, but rather linked by
  the equation (\ref{B1eq2A3}). However such considerations were done under the
  assumption that the shear tensor vanishes locally, which is never the case in
  the Bjorken picture unless all the dissipative contributions are negligible.}
it is reassuring that at least in this case the gravity picture is capable of
capturing the ambiguities of 
the boundary phenomenological construction.

\subsection{A phenomenological definition of black brane entropy}

The freedom in the definition of the hydrodynamic entropy current on the gravity
side follows not from the various possible notions of horizon, but rather from
adopting the phenomenological construction in the bulk, which is analogous to
the boundary one. The surfaces considered are not horizons in any of the usual
senses, but they do have the property that their area increases, and that is all
that is required if one adopts the slowly-evolving geometry approach. In order
to understand this in greater generality one would of 
course need to go beyond the Bjorken flow example and consider the equation
\rf{AreaExp} evaluated on the gravity dual to the general hydrodynamics. This is
certainly possible employing the Weyl-covariant formulation in the bulk. It
would be very interesting see this in detail.

\section{Conclusions and outlook}

The main goal of this article was to explore the relationship between the
notions of entropy on both sides of AdS/CFT duality. This lead to a
phenomenological definition of black brane entropy on the gravity side, which
was inspired 
by the corresponding construction in hydrodynamics. In the case of Bjorken flow
the freedom inherent in this definition accounts for the entire ambiguity
appearing in the hydrodynamic entropy current in this case. This leads to an
understanding why the event horizon coincides in the leading and first
non-leading orders of the gradient expansion with the unique apparent horizon
compatible with the boundary flow. The origin of this circumstance is the Weyl
invariance of the boundary theory. It would be very interesting to understand the form of general hydrodynamic entropy current from the bulk perspective and carefully understand the sources of possible freedom in such definition.

It is natural to ask what is the physical relevance of the potential ambiguity in the definition of entropy current. In the
case of local entropy production such an ambiguity signals 
lack of physical meaning. This however should not be disturbing, because the
thermodynamic notion of entropy makes sense only in equilibrium. Since one
expects that systems described by hydrodynamics 
equilibrate due to dissipative effects, the total
entropy can be calculated in the late stages of evolution and is given by the
thermodynamic entropy. On the gravity side this translates to the
notion of isolated horizons as those for which entropy can be defined
precisely. The framework of slowly evolving 
geometry is especially useful here, since it quantifies how far from equilibrium
the black brane is by examining the 
validity of the first law of thermodynamics. This tool can be used as a probe of
when and where the strongly coupled boundary quantum field theory is close to
equilibrium.

Apart from some more or less obvious generalizations it would be interesting to
explore these ideas in the context of equilibration of the boundary quantum
field theory perturbed out of equilibrium by localized sources (in the spirit of
\cite{Chesler:2008hg,Chesler:2009cy,Bhattacharyya:2009uu}). This application of
the slowly evolving horizons framework is especially interesting because in the
case of planar horizons considered here there can be widely separated regions,
of which some are in local equilibrium while others are not.

On the gravity side these ideas also raise some interesting
questions. Throughout this paper we have often referred to various surfaces
being ``close'' to a future outer trapping horizon (FOTH). In practice this has
always meant close along surfaces 
of constant Eddington-Finkelstein coordinate $v$ in terms of the radial
coordinate $r$ of the adapted coordinate 
system. From a purely geometric point of view this is not really satisfactory;
notions of distance shouldn't be formulated in terms of an arbitrary coordinate
system. Thus, more thought needs to be given to properly understanding whether
there is an invariant way to define this closeness. One such idea has been
addressed in \cite{BoothMartin} where the authors considered the minimum proper
time interval between a spacelike FOTH and a null event horizon. However the
results are not entirely satisfactory and work remains to be done. Ideally one
would like to obtain a general proof showing that in the vicinity of a slowly
evolving horizon there would always be candidate event and alternate horizons
which are arbitrarily close in a well-defined manner. How to do this is
currently an open problem.

This work also sheds some light on the long-standing discussion as to whether it
is more ``correct'' to consider event or apparent horizons in physical
situations. Instead of there being just two choices, we now propose that are
many more and the uncertainty as to which one should considered actually
reflects a physical ambiguity in the proper definition of entropy and other
quantities such as energy as one moves away from equilibrium. Again these issues
deserve further investigation.

\acknowledgments
The authors would like to thank Romuald Janik, Rob Myers and Paul Romatschke for valuable discussions. During the course of this work, MH has been
supported by Polish Ministry of Science and Higher Education grants N N202 247135 and N N202 105136 and the Foundation 
for Polish Science award START. The authors acknowledge the use of Matthew Headrick's excellent Mathematica package {\tt diffgeo.m} for symbolic GR calculations \cite{Headrick}.

\begin{appendix}

\section{Derivation of the constraint equations}
\label{appA}

In this appendix we consider the derivations of the geometric constraints
mentioned at the end of section \ref{geometry}. A more complete discussion of
almost all of the results mentioned in this section can be found in
\cite{Booth:2006bn}.

To begin we write $\mathcal{V}^a$ in the more general form
\bea
\mathcal{V}^a = A \ell^a - C n^a \, ,
\eea
which will help to display symmetries in the equations. In the main text $A$ is set back to unity.

Now, in finding the equations constraining the derivatives of the extrinsic
geometry the key piece of information that must be  
used is that $\ell_a$ and $n_a$ are always normal to the $S_\lambda$. Thus 
\bea
\tq_a^b \Lie_\cV \ell_b  =  0  \; & \Longrightarrow  \; \cV^b \nabla_b \ell_a  = &  - ({d}_a  - \tom_a) C + \kappa_\cV \ell_a  \; \label{Xnab} \mbox{and} \\
  \tq_a^b \Lie_\cV n_b  = 0 \;  & \Longrightarrow \; \cV^b \nabla_b n_a  = &   (d_a +\tom_a) A   - \kappa_\cV n_a \, ,  \label{Xnab2} 
\eea
The importance of these two relations is that they can be used to replace covariant derivatives in the $\mathcal{V}^a$ direction with 
derivatives in the tangent direction plus a part proportional to $\kappa_\cV$:
the $\cV^a$ component of the connection on the normal bundle.  

It is then possible to apply these relations to find expressions for the rate of
change of components of the extrinsic geometry. For $\tl$ one 
proceeds in the following manner. First, 
\bea
\Lie_\cV \tl = \cV^c \nabla_c \tl = (\cV^c \nabla_c \tq^{ab}) \nabla_a \ell_b + \tq^{ab}  \cV^c \nabla_c (\nabla_a \ell_b) \, . \label{E0}
\eea
Expanding the first term with (\ref{SurfMet}) and then applying (\ref{Xnab}) and (\ref{Xnab2}) gives
\bea
(\cV^c \nabla_c \tq^{ab}) \nabla_a \ell_b & = & (d^b A + \tom^b A) (\ell^a \nabla_a \ell_b) 
- (d^b C - \tom^b C) (n^a \nabla_a \ell_b)  \label{E1}  \\ 
& & +  (d^b C - \tom^b C) \tom_b  \nn \, , 
\eea
while for the second term, the Riemann tensor can be used to commute the indices of the double derivative to obtain
\bea
\tq^{ab} \cV^c \nabla_c \nabla_a \ell_b  = \tq^{ab} \cV^c (\mathcal{R}_{cabd} \ell^d + \nabla_a \nabla_c \ell_b) \, . \label{E2}
\eea
Next, applying the definition of the Einstein tensor $G_{ab} = \mathcal{R}_{ab} - \frac{1}{2} \mathcal{R} g_{ab}$ along with the
Gauss relation
\bea
\tq_a^e \tq_b^f \tq_c^g \tq_d^h \mathcal{R}_{efgh} 
= \tilde{R}_{abcd}
+ (k^{(\ell)}_{ ac}  k^{(n)}_{ bd} + k^{(n)}_{ ac}  k^{(\ell)}_{ bd}) 
-  (k^{(\ell)}_{ bc}  k^{(n)}_{ ad} + k^{(n)}_{ bc}  k^{(\ell)}_{ ad}) \, \, , \label{Gauss}
\eea
where $\tilde{R}_{abcd}$  is  Riemann tensor associated with the $(n-1)$-dimensional $\tq_{ab}$, 
the Riemann term of (\ref{E2}) becomes
\bea
\tq^{ab} \cV^c\mathcal{R}_{cabd} \ell^d = - A G_{ab} \ell^a \ell^b  - C G_{ab} \ell^a n^b + \frac{1}{2} \tilde{R} + \tl \tn 
- k^{(\ell)}_{ab} k^{ab}_{(n)} \, . \label{Ea} 
\eea
Meanwhile, again with the help of (\ref{Xnab}) the other term is
\bea
\tq^{ab} \cV^c \nabla_a \nabla_c \ell_b &=& - d^2 C + d_a (C \tom^a) + \kappa_{\cV} \tl - k_{ab}^{(\cV)} k^{ab}_{(\ell)} \\
& & - (d^b A + \tom^b A) (\ell^a \nabla_a \ell_b) + (d^b C - \tom^b C) (n^a \nabla_a \ell_b) \nn \, . 
\eea

Combining these results there is some cancellation and so one gets
\bea
\Lie_{\cV} \tl &=& \kappa_{\cV} \tl - d^2 C + 2 \tom^a d_a C - C \left[ \norm \tom \norm^2 - d_a \tom^a - \tilde{R}/2 + G_{ab} \ell^a n^b - \tl \tn \right] \nn \\
& & - A \left[ k^{(\ell)}_{ab} k_{(\ell)}^{ab} + G_{ab} \ell^a \ell^b  \right] \, ,
\eea
where $\| \tom \|^2 = \tom_a \tom^a$.
In this form the relation is independent of the spacetime dimension, however, if one decomposes the extrinsic curvature term 
on the second line with (\ref{kDecomp}) the dimension appears
\bea
\Lie_{\cV} \tl &=& \kappa_{\cV} \tl - d^2 C + 2 \tom^a d_a C - C \left[\norm
  \tom \norm^2 - d_a \tom^a - \tilde{R}/2 + G_{ab} \ell^a n^b - \tl \tn \right]
\nn \\ 
& & - A \left[ \frac{\tl^2}{n-1}   + \norm\sigma_{(\ell)}\norm^2 + G_{ab} \ell^a \ell^b \right] \, , \label{explderiv2}
\eea
where and  $\| \sigma^{(\ell)} \|^2 = \sigma^{(\ell)}_{ab} \sigma_{(\ell)}^{
  ab}$. Note in particular that the only quantity not directly defined on $S_v$
is the gauge-dependent $\kappa_{\cV}$. 

The rate of change of the inward expansion may be obtained simply by exchanging
$\ell$ and $n$ and swapping $A \leftrightarrow - C$ in 
(\ref{explderiv}). Then
\bea
  \Lie_\cV \theta_{(n)}  &=& -  \kappa_\cV \tn + d^2 A + 2 \tom^a d_a A + 
   A \left[\norm\tilde{\omega}\norm^2  + {d}_{a} \tilde{\omega}^{a} - \tilde{R}/2 
    + G_{ab} n^{a} \ell^{b} - \tl \tn \right] \nn
     \\
  && \quad 
  + C \left[  \frac{\theta_{(n)}^{2} }{n-1}+ \norm \sigma^{(n)}\norm^{2} 
    + G_{ab} n^{a} n^{b} 
     \right] \,   \label{expnderiv2}
      \, .
\eea

A similar calculation using (\ref{Xnab}) and (\ref{Xnab2}) finds the derivatives of the expansions pulled back onto the $S_\lambda$. 
First again using the Riemann tensor to commute derivatives one can show the Codazzi relations
\bea
\tq_a^e \tq_b^f \tq_c^g \ell^h \mathcal{R}_{efgh} 
&=& (d_a - \tom_a) k^{(\ell)}_{ bc} 
- (d_b - \tom_b) k^{(\ell)}_{ ac} \label{Codazzi} \,  \\
\tq_a^e \tq_b^f \tq_c^g n^h \mathcal{R}_{efgh} 
&=& (d_a + \tom_a) k^{(n)}_{ bc} 
- (d_b + \tom_b) k^{(n)}_{ ac} \nn \, . 
\eea
Now, the extrinsic curvatures can be decomposed into expansions and shears and while the Riemann tensor can be 
decomposed into Weyl and Ricci components 
using the well-known relation  \cite{Hawking:1973uf}:
\bea
\mathcal{R}_{abcd} = \mathcal{C}_{abcd} 
+  \frac{2}{n-1} \left( g_{a [ c} \mathcal{R}_{d] b} + g_{b [ d} \mathcal{R}_{c] a} \right) 
- \frac{2}{n(n-1)} \left( g_{a [c} g_{d] b} \right) \, , \label{weyl} 
\eea
where $\mathcal{C}_{abcd}$ is the Weyl tensor, square brackets indicate anti-symmetrization and the reader should keep in mind
that we are considering an $(n+1)$-dimensional spacetime.  Then  constracting with $\tq^{bc}$ we find 
\begin{eqnarray}
  {d}_{a}  \theta_{(\ell)} &=& \tl \tom_a
  2 ({d}_{b} - \tom_b) \sigma^{(\ell) b}_a
  - \frac{1}{n-1} \tilde{q}_{a}^{b}  G_{bc}\ell^{c} 
    - 2 \tq_a^b  \mathcal{C}_{bcde} \ell^{c} \ell^{d} n^{e} \mbox{ and}
  \label{da_thetal2}\\
  {d}_{a}  \theta_{(n)} &=& - \tn \tom_a
  2 ({d}_{b} + \tom_b) \sigma^{(n)  b}_a 
  - \frac{1}{n-1} \tilde{q}_{a}^{b} G_{bc} n^{c} 
    + 2 \tq_a^b \mathcal{C}_{bcde} n^{c} \ell^{d} n^{e}  \, \, . 
\end{eqnarray}

The final equation of interest for this paper is the rate of change of the angular momentum one-form up $\triangle$. This comes from a 
direct expansion of  $\tq_a^b \Lie_\cV (n_c \nabla_b \ell^c)$ along with the familiar applications of the Riemann tensor to commute derivatives,
(\ref{Xnab}) and (\ref{Xnab2}) to enforce the fact that the $S_\lambda$ fit together smoothly into a surface, and then breaking the Riemann
tensor into Weyl and Ricci components. We find
\bea
    \Lie_\cV \tilde{\omega}_{a}    &=&  {d}_a \kappa_\cV -  k^{(\ell)}_{ a b}
   \left[ {d}^b A  +  \tilde{\omega}^b A \right]
  - k^{(n)}_{ab}
  \left[{d}^b C - \tilde{\omega}^b C \label{dkappa2}
    \right] \\
 & &   + \tilde{q}_{a}^{\; \; b} \left[ \frac{1}{(n-1)} G_{bc} (\ell^c + C n^c)     - \mathcal{C}_{bcde} {\cV}^{c}\ell^{d}n^{e} 
\right]    \nonumber \, , 
 \end{eqnarray}

As shown in the main text, these equations lie at the root of black hole mechanics and dynamics.

\section{Bjorken flow at second order}
\label{appb}

The second order solution \cite{Heller:2009zz,Kinoshita:2008dq} for the
coefficients in \rf{efmetric} reads  
\bea
A_{2}\left( v \right) &=& -\frac{2 \pi ^2 \log (v) \Lambda ^2}{3
v^4}+\frac{\pi ^2 \log \left(\frac{v^2}{\pi ^2 \Lambda ^2}+1\right)
\Lambda^2}{3 v^4}+
 \frac{2 \pi ^2 \log (\pi  \Lambda ) \Lambda ^2}{3
v^4}-\\ \nonumber
&-&\frac{\pi ^2 \log (2) \Lambda ^2}{9
  v^4}+\frac{\left(v^4-3 \pi ^4 \Lambda ^4\right) \delta _1^2}{9
v^6}-\frac{-2 v^4+\pi ^2 \Lambda ^2 v^2+4 \pi ^3
  \Lambda ^3 v + 
2 \pi ^4 \Lambda ^4}{6 v^6} + \\ \nonumber 
&+& \frac{2 \left(3 v^4-2 \pi^3 \Lambda ^3 v-3 \pi ^4 \Lambda ^4\right)
  \delta _1}{9 v^6}-\frac{2 \left(v^4+\pi ^4 \Lambda ^4\right) \delta_2}{3 
  v^5}+ \\ \nonumber
&+& \frac{\left(v^4+\pi ^4 \Lambda
  ^4\right) \tan ^{-1}\left(\frac{v}{\pi  \Lambda }\right)}{3 \pi
v^5 \Lambda }
\\
b_{2}' \left( v \right) &=& \frac{8 \log (v) v^2}{9 \pi ^5 \Lambda ^5-9
\pi  v^4 \Lambda }-\frac{8 \log (\pi  \Lambda ) v^2}{9 \pi ^5 \Lambda
  ^5-9 \pi  v^4 \Lambda }-\frac{\left(7 v^4-4 \pi ^3 \Lambda ^3 v-3
\pi ^4 \Lambda ^4\right) \tan
  ^{-1}\left(\frac{v}{\pi  \Lambda }\right)}{9 \pi  \Lambda
\left(v^6-\pi ^4 v^2 \Lambda ^4\right)}+\\ \nonumber
&+& \frac{4 \pi
  ^2 \Lambda ^2 \log (2)}{9 \pi ^4 v \Lambda ^4-9 v^5}+\frac{2
\left(v^2+\pi  \Lambda  v+\pi ^2 \Lambda
  ^2\right)}{9 \Lambda  (v+\pi  \Lambda ) \left(v^2+\pi ^2 \Lambda
^2\right) v}-\frac{2 \left(v^2-\pi  \Lambda
  v+\pi ^2 \Lambda ^2\right) \log \left(\frac{v^2}{\pi ^2 \Lambda
^2}+1\right)}{9 \pi  \Lambda  (\pi  \Lambda -v)
  \left(v^2+\pi ^2 \Lambda ^2\right) v}+ \\ \nonumber
&+& \frac{4 \left(v^2+\pi
\Lambda  v+\pi ^2 \Lambda ^2\right) \log
  \left(\frac{v}{\pi  \Lambda }+1\right)}{9 \pi  \Lambda  (v+\pi
\Lambda ) \left(v^2+\pi ^2 \Lambda ^2\right)
  v}+\frac{2 \delta _2}{3 v^2}+\frac{2 \delta _1^2}{9 v^3}-  \\ \nonumber 
&-&\frac{11
v^6+22 \pi  \Lambda  v^5+34 \pi ^2 \Lambda ^2
  v^4+60 \pi ^3 \Lambda ^3 v^3+43 \pi ^4 \Lambda ^4 v^2+26 \pi ^5
\Lambda ^5 v+16 \pi ^6 \Lambda ^6}{9 (v+\pi
  \Lambda )^2 \left(v^2+\pi ^2 \Lambda ^2\right)^2 v^3}+ \\ \nonumber 
&+&\frac{4
\left(3 v^6+6 \pi  \Lambda  v^5+9 \pi ^2 \Lambda
  ^2 v^4+7 \pi ^3 \Lambda ^3 v^3+5 \pi ^4 \Lambda ^4 v^2+3 \pi ^5
\Lambda ^5 v+\pi ^6 \Lambda ^6\right) \delta
  _1}{9 (v+\pi  \Lambda )^2 \left(v^2+\pi ^2 \Lambda ^2\right)^2 v^3}
\\
c_{2} \left( v \right) &=& d_{2} \left( v \right) - b_{2} \left( v \right)
\\ 
d_{2} \left( v \right) &=&  -\frac{\delta _1^2}{6 v^2}-\frac{\delta
_1}{3 v^2}-\frac{v-\pi  \Lambda }{3 \pi  v^2 \Lambda }-\frac{(v-3 \pi
  \Lambda ) \tan ^{-1}\left(\frac{v}{\pi  \Lambda }\right)}{6 \pi ^2
v \Lambda ^2}-\frac{2 \log (v)}{3 \pi ^2
  \Lambda ^2}+ \\ \nonumber 
&+& \frac{\log \left(\frac{v^2}{\pi ^2 \Lambda
^2}+1\right)}{4 \pi ^2 \Lambda ^2}+\frac{\log
  \left(\frac{v}{\pi  \Lambda }+1\right)}{6 \pi ^2 \Lambda
^2}+\frac{2 \log (\pi  \Lambda )}{3 \pi ^2 \Lambda
  ^2}-\frac{\delta _2}{v}+\frac{1}{12 \pi  \Lambda ^2}
\eea
The expression for $b_2'$ given above can be integrated in terms of
polylogarithmic functions, but the explicit form of the integral will not
be needed in the sequel. The quantities $\delta_{1}$ (and $\delta_{2, 3,
  \ldots}$ in higher orders) are integration constants related to residual gauge
symmetry $r \rightarrow r + R\left( \tilde{\tau} \right)$ mentioned earlier.

\end{appendix}

\end{document}